\documentclass[12pt]{article}
\pdfoutput=1
\usepackage{latexsym, graphicx} 
\usepackage{amsmath}
\usepackage{amssymb}

 \newcommand{\arXiv}[1]{\href{http://www.arXiv.org/abs/#1}{#1}}
\usepackage[colorlinks=true, linkcolor=blue, bookmarks=true]{hyperref}

\makeatletter
\renewcommand\section{\@startsection {section}{1}{\z@}%
                                   {-3.5ex \@plus -1ex \@minus -.2ex}
                                   {2.3ex \@plus.2ex}%
                                   {\normalfont\large\bfseries}}
\renewcommand\subsection{\@startsection{subsection}{2}{\z@}%
                                     {-3.25ex\@plus -1ex \@minus -.2ex}%
                                     {1.5ex \@plus .2ex}%
                                     {\normalfont\bfseries}}
\makeatother
\newcommand{\beq}{\begin{equation}}
\newcommand{\eeq}{\end{equation}}
\newcommand{\ber}{\begin{array}}
\newcommand{\eer}{\end{array}}

\newcommand{\dsty}{\displaystyle}

\newcommand{\te}{\theta}

\newcommand{\de}{\delta}

\newcommand{\eps}{\varepsilon}
\newcommand{\ups}{\upsilon}
\newcommand{\om}{\omega}

\newcommand{\ena}{\end{eqnarray}}
\newcommand{\beqa}{\begin{eqnarray}}
\newcommand{\eeqa}{\end{eqnarray}}
\newcommand{\bea}{\begin{eqnarray}}
\newcommand{\eea}{\end{eqnarray}}

\newcommand{\be}{\begin{equation}}
\newcommand{\ee}{\end{equation}}

\begin{document}
\begin{titlepage}
\begin{flushright}
\phantom{arXiv:yymm.nnnn}
\end{flushright}
\vfill
\begin{center}
{\Large\bf Renormalization group,
secular term resummation and AdS (in)stability}    \\
\vskip 15mm
{\large Ben Craps$^{a,b}$, Oleg Evnin$^c$ and Joris Vanhoof$^a$}
\vskip 7mm
{\em $^a$ Theoretische Natuurkunde, Vrije Universiteit Brussel and\\
The International Solvay Institutes\\ Pleinlaan 2, B-1050 Brussels, Belgium}
\vskip 3mm
{\em $^b$ Laboratoire de Physique Th\'eorique, Ecole Normale Sup\'erieure,\\ 24 rue Lhomond, F-75231 Paris Cedex 05, France}
\vskip 3mm
{\em $^c$ Department of Physics, Faculty of Science, Chulalongkorn University,\\
Thanon Phayathai, Pathumwan, Bangkok 10330, Thailand}

\vskip 3mm
{\small\noindent  {\tt Ben.Craps@vub.ac.be, oleg.evnin@gmail.com, Joris.Vanhoof@vub.ac.be}}
\vskip 10mm
\end{center}
\vfill

\begin{center}
{\bf ABSTRACT}\vspace{3mm}
\end{center}

We revisit the issues of non-linear AdS stability, its relation to growing (secular) terms in na\"\i ve perturbation theory around the AdS background, and the need and possible strategies for resumming such terms. To this end, we review a powerful and elegant resummation method, which is mathematically identical to the standard renormalization group treatment of ultraviolet divergences in perturbative quantum field theory. We apply this method to non-linear gravitational perturbation theory in the AdS background at first non-trivial order and display the detailed structure of the emerging renormalization flow equations. We prove, in particular, that a majority of secular terms (and the corresponding terms in the renormalization flow equations) that could be present on general grounds given the spectrum of frequencies of linear AdS perturbations, do not in fact arise.

\vfill

\end{titlepage}

\section{Introduction}

Time-dependent perturbation theory is often plagued by secular terms. While suppressed by the expansion parameter, secular terms grow with time. They invalidate the na\"\i ve perturbation theory at time scales that are typically proportional to some inverse power of the expansion parameter. In order to extend the validity of perturbation theory to larger times, secular terms need to be resummed.

To set the stage, consider the anharmonic oscillator \cite{landau,logan}
\be
\ddot x+x+\epsilon x^3=0
\ee
with small positive $\epsilon$, and construct a perturbative solution $x(t)=x_0(t)+\epsilon x_1(t)+\cdots$. For initial conditions $x(0)=1,\ \dot x(0)=0$, the zeroth order solution is $x_0(t)=\cos t$. At first order, we find
\be
\ddot x_1 + x_1 = -\frac{1}{4}\cos 3t-\frac{3}{4}\cos t.
\ee
The last term is a resonant source term, giving rise to a secular term proportional to $t\sin t$:
\be
x(t)=\cos(t)+\epsilon\left[\frac{1}{32}(\cos 3t-\cos t )-\frac{3}{8}\,t\sin t \right]+\cdots
\label{oscnaive}
\ee
According to the Poincar\'e-Lindstedt method, the secular term can be absorbed in a small frequency shift, leading to a generalized asymptotic expansion that provides more accurate approximations for longer time intervals,
\be
x(t)=\cos\left[\left(1+\frac{3\epsilon}{8}+\cdots\right)t\right]+\frac{\epsilon}{32}\cos\left[3\left(1+\frac{3\epsilon}{8}+\cdots\right)t\right]+\cdots
\label{oscresum}
\ee 

For a single oscillator described by a Hamiltonian, this is all that is needed. When dealing with dissipative systems, or with multiple oscillators exhibiting resonances, one needs more elaborate resummation techniques, known as multiple scale methods. A particularly elegant such method was proposed by Chen, Goldenfeld and Oono \cite{Chen:1995ena}, and is based on the renormalization group (RG). We will review it in section~\ref{sec:RG} and rely on it in the remainder of this paper. 

Our focus will be on Hamiltonian systems. Historically, the study of perturbation theory was driven by celestial mechanics, in particular the question whether the solar system is stable on very long time scales (given that interactions between planets perturb the Keplerian orbits). These studies culminated in the Kolmogorov-Arnol'd-Moser (KAM) theory \cite{KAM}, which showed with mathematical rigor that both stable and unstable orbits exist, depending on whether unperturbed frequencies are resonant. The stable orbits occur for non-resonant frequencies, and correspond to small perturbations of the unperturbed orbits.

In recent years, secular terms have appeared quite prominently in studies of non-linear stability of anti-de Sitter (AdS) space and closely related spacetimes. In \cite{Bizon:2011gg}, Bizo\'n and Rostworowski provided numerical evidence that arbitrarily weak spherically symmetric perturbations can cause global AdS to collapse into a black hole (possibly after multiple scatterings from the AdS boundary). In addition, they showed that in weakly nonlinear perturbation theory secular terms appear that cannot be removed by frequency shifts, and suggested that these secular terms signal a turbulent flow of energy to higher and higher frequencies. These additional secular terms (beyond those that can be removed via the Poincar\'e-Lindstedt method) arise from resonances in the spectrum of a scalar field in global AdS, as we will review in section~\ref{sec:AdS}. Many papers have further investigated this and related systems, mostly using numerical general relativity supplemented with weakly nonlinear perturbation theory (see, for instance, \cite{Dias:2011ss,Dias:2012tq,Maliborski:2013jca,Buchel:2013uba,Abajo-Arrastia:2014fma,Maliborski:2014rma,Balasubramanian:2014cja}). After quite a few surprises, a rich phenomenology has been uncovered, with the space of initial conditions exhibiting islands of stability within a sea of instability.

These results raise several conceptual questions: Can the additional terms be removed by multiple scale/RG techniques? What is the precise relation between AdS instability and secular terms? Does the weakly nonlinear perturbation theory exhibit additional structure that can be uncovered by analytical means?

Some of these questions were also addressed in the recent paper \cite{Balasubramanian:2014cja}, in which a multiple scale method (referred to as ``Two Time Framework'' and valid to first non-trivial order in the perturbation) was applied to the system of \cite{Bizon:2011gg} truncated to a finite set of modes. The resulting equations were studied numerically, and the output was compared with results from numerical general relativity. One point that was emphasized in \cite{Balasubramanian:2014cja} is that secular terms that cannot be absorbed in frequency shifts do not necessarily imply AdS instability. Due to the finite time range of numerical simulations and/or the required resolution, it is not always straightforward, however, to reach firm conclusions on the long-time behavior of given initial conditions; for instance, the fate of certain ``two-mode'' initial data discussed in \cite{Bizon:2011gg} and \cite{Balasubramanian:2014cja} is still being debated.\footnote{We thank P.~Bizo\'n, A.~Buchel and L.~Lehner for correspondence on this issue.} This illustrates the fact that a more systematic understanding would be very welcome.

The purpose of the present work is to resum systematically the secular terms of \cite{Bizon:2011gg} using the RG method of  \cite{Chen:1995ena}.\footnote{In the context of the AdS/CFT correspondence, the RG method of \cite{Chen:1995ena} has recently also appeared in \cite{Nakayama:2013fha, Kuperstein:2013hqa}.}  Our RG setup agrees to lowest non-trivial order with the Two Time Framework of \cite{Balasubramanian:2014cja}, but the focus is different. Our work will be analytic rather than numerical, leading to explicit results at first order that are valid for all modes. In particular, we will show analytically that a majority of secular terms that could be present on the basis of frequency relations among the linearized AdS perturbations are in fact absent, and will provide explicit analytic expressions for all nonvanishing secular terms. Note that deriving all-mode expressions for the secular term coefficients is not a matter of pure pedantry. In a system prone to turbulence, one expects that high frequency modes typically get involved in the evolution. Having all-mode expressions for the secular term coefficients (and the corresponding energy drift), and in particular their ultraviolet asymptotics, is likely to be crucial for any analytic considerations of the turbulent behavior.

It is interesting that weakly nonlinear perturbation theory in global AdS exhibits a fully resonant spectrum (which drives instability), but at the same time a majority of secular terms allowed for such a spectrum are in fact absent (which weakens the instability). This interplay of conflicting factors may underlie the apparent complexity of the AdS stability domain that has been observed numerically \cite{Bizon:2011gg,Dias:2011ss,Dias:2012tq,Maliborski:2013jca,Buchel:2013uba,Abajo-Arrastia:2014fma,Maliborski:2014rma,Balasubramanian:2014cja}. We provide some comments on relations between the absence of some classes of secular terms and the abundance of quasi-periodic solutions in section \ref{sec:qp}.
  
The general stucture of the paper is as follows: Section~\ref{sec:RG} contains a systematic discussion of the RG method introduced in \cite{Chen:1995ena}. In section~\ref{sec:AdS} this method is applied to weakly nonlinear perturbation theory in AdS. Two appendices contain technical details on our computations.


\section{Renormalization group resummation}\label{sec:RG}

\subsection{Ubiquity of secular terms}

When dealing with a system subject to a small perturbation, it is natural to describe its evolution by an asymptotic series in the perturbation magnitude, an approach familar under the name of perturbation theory. This strategy is equally applicable when one perturbs the initial conditions rather than the definition of the system proper. Solutions are then presented as an asymptotic series in the magnitude of the deviation from the specific chosen initial conditions, for which an exact solution is know.

The nature and accuracy of such an asymptotic series approximation is necessarily subtle, except for the rare cases when the asymptotic series happens to converge. Nonetheless, the usual practical wisdom tells us that, as long as the subsequent terms in the series are smaller than the preceding ones, the expansion is usable and sound. In fact, some of the most precise predictions in physics have been made using such truncated asymptotic expansions (a very slippery step from a purely mathematical perspective).

If one fixes the time interval whereupon the evolution is considered, and diminishes the magnitude of the perturbation, the higher-order terms in the asymptotic expansion diminish relatively to the
lower-order terms (being weighted by higher powers of the perturbation magnitude). One is then in a regime when the asymptotic series is expected to approximate the exact evolution on the said fixed time interval more and more accurately.

Unfortunately, evolution due to a small perturbation on a fixed time interval is usually not what a physicist wants to consider. A problem of much greater phenomenological significance is to be able to trace the effect of a small perturbation over large times, when its impact on the evolution becomes appreciable despite its smallness. This is precisely the regime when the so-called secular terms in perturbation theory come into play.

Prototype examples of this sort come from celestial mechanics. The solar system is, to a high degree of accuracy, an integrable system described by the planets moving in the central potential of the Sun, whose position is fixed at the origin. Yet, interplanetary interactions and other physical processes (including processes of non-gravitational nature), introduce small perturbations to the idealized integrable picture. Jupiter, in particular, exerts a relatively strong influence on the motion of the Earth. The physical question is not in the minuscule corrections such perturbations induce over, say, one revolution of the Earth around the Sun, but rather how such minuscule corrections accumulate over a large time to produce substantial effects. This is precisely the question that na\"\i ve perturbation theory fails to answer.

Indeed, as described above, the magnitude of higher-order terms in  na\"\i ve perturbation theory is guaranteed to decrease on a fixed time interval when the magnitude of the perturbation is decreased, but nothing prevents a growth of the coefficients of higher orders in  na\"\i ve perturbation theory. This growth, if present, will make the asymptotic series unusable at large times, as higher-order terms will be comparable in magnitude to lower-order terms. In fact, the growing terms at higher orders in the  na\"\i ve perturbation theory typically appear in realistic situations, and they have become known as `secular' terms (from the Latin word for `century', referring to terms that become significant when considering planetary perturbations over the course of centuries). Such terms need to be restructured by means of resummation, if one is aiming at a perturbative description of the large-time dynamics at all.

Perhaps the easiest way to appreciate the ubiquity of secular terms is to examine them in a quantum-mechanical setting. The linearity of the Schr\"odinger equation allows one to write an explicit result for all orders of the perturbative expansion. Since each classical system is a limit of the corresponding quantum system, the presence of secular terms in the quantum formalism sheds some light on classical Hamiltonian systems as well.

Consider for a moment a general perturbed quantum system described by the Hamiltonian, $H=H_0(t)+\lambda V(t)$ and the corresponding evolution operator $U(t)$ satisfying
\beq
i\frac{dU}{dt}=HU,\qquad U(0)=1.
\eeq
Converting to the interaction picture, we introduce $u(t)=U_0^\dagger(t)U(t)$, where $U_0(t)$ satisfies
\beq
i\frac{dU_0}{dt}=H_0U_0,\qquad U_0(0)=1.
\eeq
Then,
\beq
i\frac{du}{dt}=\lambda {\ups}(t)u, \qquad u(0)=1,
\label{int}
\eeq
with $\ups(t)=U_0^\dagger(t)V(t)U_0(t)$.
For (\ref{int}), one obtains the standard na\"\i ve perturbative expansion
\beq
u(t)=1-i\lambda\int\limits_0^t dt_1\ups(t_1)+(-i\lambda)^2\int\limits_0^t dt_1\int\limits_0^{t_1}dt_2\ups(t_1)\ups(t_2)+
\cdots
\label{naive}
\eeq
This expansion is generically plagued by secular terms at large $t$, except for special situations like scattering, when (perhaps for a subset of matrix elements) $\ups(t)$ effectively vanishes outside a finite time interval. Indeed, unless the interactions are effectively cut off in this fashion, the natural scale of the $n$th order term in (\ref{naive}) is $\lambda^nt^n$ rather than simply $\lambda^n$, which means that the na\"\i ve perturbative expansion becomes completely useless at $t\sim 1/\lambda$. (The details, of course, depend on the particular time dependences involved.)

For the familiar case of time-independent $H_0$ and $V$, it is well-known from textbooks that (\ref{naive}) is not the right way to expand. Indeed, in the standard approach (which occasionally goes under the name of Rayleigh-Schr\"odinger perturbation theory) one expands the eigenstates and their energies in a power series in $\lambda$, rather than expanding the evolution operator. Since the energies enter the evolution operator through $\exp(-iE_nt)$, correcting the unperturbed energies by a power series in $\lambda$ is analogous to shifting the oscillator frequencies in (\ref{oscresum}). In fact, the Rayleigh-Schr\"odinger perturbation theory can be derived from (\ref{naive}) by a resummation analogous to the one leading from (\ref{oscnaive}) to (\ref{oscresum}).

Our purpose for the rest of this section will be to review some approaches to secular term resummation in a general setting, before returning to the case of non-linear gravitational perturbation theory in the AdS background in section \ref{sec:AdS}.

\subsection{Frequency adjustment and multi-scale resummation}

We shall now examine the question of what kind of secular terms may arise when a particular perturbation is applied in the context of classical Hamiltonian systems. It will be sufficient for us to focus on linear unperturbed systems with perturbations polynomial in the canonical variables. For one thing, our main goal in this paper is to shed some light on the dynamics of weakly non-linear gravitational perturbations in the AdS background. In this context, the unperturbed system is linearized gravity in the AdS background. Similar set-ups will be produced by other weakly non-linear perturbative expansions. (More generally, when dealing with the effect of an explicit dynamical perturbation on a given solution of a non-linear system, one can always treat non-linearities, expanded in the vicinity of that given solution, as merely an additional contribution to the perturbation.) As to the restriction to polynomial perturbations, it is also natural in the context of studying small deviations from a given exact solution, since, at any given order of the perturbation theory, a non-polynomial pertubation can be identically replaced by its truncated polynomial expansion up to this order.

If the unperturbed system is linear, one can always switch to the normal coordinates $c_i$, for which the unperturbed solutions are simply
\beq
c^{(0)}_i(t)=a_i\cos\te_i(t),\qquad \te_i(t)=\om_i t+ b_i.
\label{nonlin}
\eeq
The exact solutions satisfy
\beq
\ddot c_i+\om_i^2 c_i = S_i,\qquad c_i(t)=c^{(0)}_i(t)+\epsilon c^{(1)}_i(t)+\epsilon^2 c^{(2)}_i(t)+\cdots,
\label{linsol}
\eeq
where $S_i$ collectively represents all the non-linear terms contributing to the equation for $c_i$.

We can now solve (\ref{linsol}) iteratively, determining each $c^{(n)}_i$ in terms of lower order corrections,
\beq
\ddot c^{(n)}_i+\om_i^2 c^{(n)}_i = S^{(n)}_i(c^{(0)},c^{(1)},\cdots,c^{(n-1)}).
\label{iter}
\eeq
What kind of terms can emerge on the right-hand side of (\ref{iter})? We may examine this question starting from the lowest order and working all the way up. $c^{(0)}$ is a pure cosine. Multiplication of $c^{(0)}$ is governed by the formula
\be
\cos\te_i\,\cos\te_j=\frac12[\cos(\te_i+\te_j)+\cos(\te_i-\te_j)].
\label{cosmult}
\ee
Then, in a general polynomial expression made of $c^{(0)}$, all the terms will be of the form
\be
\cos(\te_{i_1}\pm\te_{i_2}\pm\te_{i_3}\pm\cdots)=\cos((\om_{i_1}\pm\om_{i_2}\pm\om_{i_3}\pm\cdots)t+ b_{i_1}\pm b_{i_2}\pm b_{i_3}\pm\cdots),
\label{sumdiff}
\ee
where $(i_1,i_2,i_3,\cdots)$ can be any set of mode numbers, and all the choices of plus and minus signs on the left hand side of (\ref{sumdiff}) are independent of each other. The right-hand side of (\ref{iter}) is a sum of such terms. What does one get for $c^{(1)}$?

If there is a contribution to $S^{(1)}_i$ of the form (\ref{sumdiff}) with a particular set of $(i_1,i_2,i_3,\cdots)$, a particular assignment for each $\pm$, and $\pm\om_i\ne\om_{i_1}\pm\om_{i_2}\pm\om_{i_3}\pm\cdots$, then this term will simply give a contribution to $c^{(1)}$ that is itself proportional to (\ref{sumdiff}). One can then safely proceed to the next order, multiplying the different contributions to $c^{(0)}$ and $c^{(1)}$ using (\ref{cosmult}) to obtain the different terms in $S^{(2)}$, all of which will again be of the form (\ref{sumdiff}), and so on ad infinitum.

The only point where this picture fails is that there may be terms with $\pm\om_i=\om_{i_1}\pm\om_{i_2}\pm\om_{i_3}\pm\cdots$. For those, substituting them to the right-hand side of (\ref{iter}) does not produce a contribution to $c^{(n)}$ of the form (\ref{sumdiff}), but rather of the form
\be
t\sin(\te_{i_1}\pm\te_{i_2}\pm\te_{i_3}\pm\cdots).
\label{secgen}
\ee
This is a secular term that grows with time and invalidates perturbation theory at sufficiently large $t$. Such terms must be eliminated by restructuring the perturbative expansion along the lines of the frequency shift we employed in going from (\ref{oscnaive}) to (\ref{oscresum}).

The kind of secular terms that may arise as we develop the $\epsilon$-expansion iteratively depends crucially on whether the spectrum of mode frequencies is {\em resonant}. `Resonant' in this context means that there exist sets of integers $m_i$ such that $\sum_i m_i\om_i=0$. If no such relations with non-zero $m_i$ exist, then the spectrum is called non-resonant.

For a non-resonant spectrum, there is only one way $\pm\om_i=\om_{i_1}\pm\om_{i_2}\pm\om_{i_3}\pm\cdots$ can be satisfied. Namely, the number of times $\om_i$ is present in the sum on the right-hand side with a plus sign should be one greater (or one smaller) than the number of times it is present with a minus sign, whereas for all other modes ($i_k\ne i$), the number of times they are present with a plus sign should be exactly the same as with a minus sign. Any other combination of $\om_{i_k}$ cannot equal $\om_i$ since that would have implied a resonant relation between the frequencies. Hence, if $\pm\om_i=\om_{i_1}\pm\om_{i_2}\pm\om_{i_3}\pm\cdots$, then (\ref{sumdiff}) becomes simply
\be
\cos(\om_i t+ b_i),
\label{nonressec}
\ee
and the corresponding secular term, i.e., the contribution to $c^{(n)}_i$ resulting from a term of the form (\ref{nonressec}) in $S^{(n)}_i$ of (\ref{iter}), becomes
\be
\eps^{n}{\cal A}^{(n)}_i(a)\,t\sin(\om_i t+ b_i),
\label{nonressecterm}
\ee
where ${\cal A}^{(n)}_i(a)$ is a certain polynomial made of the amplitudes $a_k$ of (\ref{nonlin}), which depends on the precise form of the non-linearities in $S$.

A key observation regarding the case of a fully non-resonant frequency spectrum is that any contribution to $c^{(n)}_i$ of the form (\ref{nonressecterm}) can be absorbed into an (amplitude-dependent) shift of the frequency $\om_i$ since
\be
\cos((\om_i+\alpha)t+ b_i)=\cos(\om_it+ b_i)-\alpha t\sin(\om_it+ b_i)+\cdots
\label{cosadj}
\ee
After the secular terms of the form (\ref{nonressecterm}) have been absorbed in this fashion, all the remaining terms in $c^{(n)}_i$ are of the form (\ref{sumdiff}), and one can proceed to order $\epsilon^{n+1}$, where the entire argument can be repeated verbatim.

We hence conclude that, for a fully non-resonant case, all secular terms can be iteratively removed from perturbation theory by perturbatively adjusting the frequencies $\om_i$. This procedure is know as the Poincar\'e-Lindstedt method. After the frequencies have been corrected, the perturbed motion is described by small corrections to the unperturbed one for longer and longer time intervals depending on the order of
accuracy in the perturbative frequency shift. (Note that the picture we have outlined is something of a pedestrian pre-requisite for the KAM theory \cite{KAM}, which takes the argument much further and develops stability arguments for the non-resonant case at finite small $\epsilon$, rather than in an unreliable framework of asymptotic expansions.)

The situation becomes more complicated when resonant relations between unperturbed frequencies are present. In that case, there may be many different addition-subtraction patterns that satisfy $\pm\om_i=\om_{i_1}\pm\om_{i_2}\pm\om_{i_3}\pm\cdots$. One then cannot specify the form of (\ref{sumdiff}) for the resonant terms beyond
\be
\cos(\om_it\pm b_{i_1}\pm b_{i_2}\pm b_{i_3}\pm\cdots)
\label{ressec}
\ee
with a (generally complicated) combination of phases. The corresponding secular term resulting from the contribution of (\ref{ressec}) to $S^{(n)}_i$ of (\ref{iter}) is
\be
\eps^{n}{\cal A}^{(n)}_i(a)\,t\sin(\om_i t\pm b_{i_1}\pm b_{i_2}\pm b_{i_3}\pm\cdots),
\label{ressecterm}
\ee
where ${\cal A}^{(n)}_i(a)$ is a certain polynomial made of the amplitudes $a_k$ of (\ref{nonlin}), which depends on the precise form of the non-linearities in $S$. 

Since the phase of the sine in (\ref{ressecterm}) does not have to equal $ b_i$, it cannot be in general absorbed into a shift of $\om_i$ by means of (\ref{cosadj}). Of course, this term could always be absorbed into a shift of {\em both} $\om_i$ and $a_i$. However, the shift of $a_i$ would have to grow linearly with time, so that one generates $t\sin(\om_i t+ b_i)$ from the shift of $\om_i$ and $t\cos(\om_i t+ b_i)$ from the shift of $a_i$, and a combination of such terms can always match (\ref{ressecterm}). By itself, a linearly growing $a_i$ is no better and no worse than the original secular term, and more powerful resummation methods are needed. We shall turn to such methods shortly.

Physically, the fact that a frequency adjustment is not sufficient for the resonant case, but one also obtains terms that look like amplitude drifts, simply means that, over large times, significant energy transfer occurs between different modes, even when the perturbation is small. This is in contrast to the non-resonant case, where a small perturbation can only induce small amplitude oscillations of the energy back and forth between different modes, without significant energy transfer occurring even over large time scales.

The long-term fate of a resonant system under the impact of a dynamical perturbation can only be determined after a resummation of secular terms has been performed. The failure of the Poincar\'e-Lindstedt frequency shift by itself should by no means be interpreted as a sign of instability. A complete resummation can produce long-period oscillations of energy between the modes, or perhaps transfer of energy to high-frequency modes (turbulence), etc. Many scenarios are possible.

There is a number of resummation methods described in the literature (see, e.g., \cite{murdock}). One encounters descriptions of the multi-scale method particularly often. We shall very quickly review this method here, only stating the general idea and referring the reader to \cite{murdock} for further details. The lowest-order multi-scale method (under the name of `Two-Time Framework') has been applied to the problem of AdS stability in \cite{Balasubramanian:2014cja}.

We have already alluded above that a general secular term can be absorbed into perturbatively small frequency adjustments and a slow drift of the amplitudes. Note that the frequency adjustment can be thought of as a slow drift of the phases. One thus arrives at the concept of absorbing secular terms into a slow variation of the integration constants in the zeroth order solution $c^{(0)}_i$. This general idea is shared by both the multiscale method and the renormalization group method we shall describe further below. Note that the Poincar\'e-Lindstedt method is a special case of this set-up, for which the phases aquire a slow linear drift, whereas the amplitudes do not evolve.

The `slow variation of the integration constants' we mentioned above is a rather vague concept and one needs to decide in practice how this dependence is distributed between different orders of perturbation theory. In the multiscale method, one introduces dependences with explicit powers of $\epsilon$ in the form 
\be
a_i=a_i(\epsilon t,\epsilon^2 t,\epsilon^3 t,\cdots),\qquad  b_i= b_i(\epsilon t,\epsilon^2 t,\epsilon^3 t,\cdots). 
\label{multscale}
\ee
When the functions are specified in this form, one knows, for example, that a term quadratic in $\epsilon$ can arise from differentiating two times with respect to the first argument or one time with respect to the second argument, etc. The term `multiscale' comes precisely from the multiple scales ($\epsilon t$, $\epsilon^2 t$, $\epsilon^3 t$, etc) involved in this construction. (At first non-trivial order, only $t$ and $\epsilon t$ are involved, hence the `Two-Time Framework' of \cite{Balasubramanian:2014cja}.)

One then substitutes (\ref{multscale}) to (\ref{nonlin}) and then to the equations of motion (\ref{iter}), and demands that the terms resulting from differentiating (\ref{multscale}) conspire in precisely such a way as to remove the resonant terms in $S_i$ (the terms whose frequency is $\om_i$). At $n$th order, this results in a differential equation that fixes the dependence of $a_i$ and $ b_i$ on $\epsilon^n t$, which is then fed to the next order. To address the issue of non-linear stability, one should examine the behavior of the amplitudes after the resummation has been performed.

The multiscale method is a powerful resummation scheme applicable in a general setting and including the Poincar\'e-Lindstedt method as a simple special case. (Whether the result of the resummation is free from pathological growth depends, of course, on the particular system at hand.) However, the need to explicitly prescribe how the slow dependence of the (unperturbed) integration constants on time is distributed between different orders of perturbation theory, as in (\ref{multscale}), may create complications in more subtle cases. For example, the case of stability analysis for Mathieu equation is mentioned in \cite{Chen:1995ena}, where an unusual scale $\epsilon^{3/2}t$ appears through resummations, and that would have to be guessed in the initial ansatz (\ref{multscale}) for the multiscale method. Even if that does not happen, the method becomes rather convoluted at higher orders, since new `slow' secular terms dependent on the slow time variables develop in (\ref{multscale}) and those `slow' secular terms have to be removed by adjusting the dependence of (\ref{multscale}) on even slower time scales. Detailed explanations can be found in \cite{murdock}. As one aspect of this highly convoluted procedure, it may turn out inconvenient, depending on the circumstances, that the multiscale method does not take as its input the na\"\i ve perturbative expansion, but rather requires re-deriving an alternative expansion order-by-order from scratch. The renormalization group method we shall present below is an alternative formulation with many appealing features.

\subsection{Renormalization group method}

In the preceding exposition, we have reviewed the general problem of secular terms in non-linear perturbation theory, the types of secular terms arising when the unperturbed system is linear (or, more generally, integrable) and Hamiltonian, and a general multiscale method for resumming such secular terms.

One could in principle rest content with this state of affairs and proceed applying the resummation techniques to our particular problem (AdS instability). However, we believe it is useful to review another resummation strategy  \cite{Chen:1995ena}, modelled on the renormalization group treatment of ultraviolet divergences in quantum field theory. This method is as powerful as multiscale resummation we have briefly reviewed above, but has the advantage in that its sole input is the na\"\i ve perturbative expansion (without the need to re-solve the perturbation theory equations). The method also has the appeal of being intuitive, especially for people with high energy theory backgrounds.

The renormalization method (just like the multiscale method we reviewed above) aims at constructing slow time dependences of the integration constants of unperturbed solutions in a way that eliminates secular terms from perturbative expansions. We shall start by a simple matter-of-fact statement of the method and applying it to secular terms of the form (\ref{ressec}), and then give some qualifying explanatory remarks. The recipe \cite{Chen:1995ena} is as follows:\vspace{2mm}

\noindent 1) Choose a moment of time $\tau$ and introduce a perturbative $\epsilon$-dependence to the integration constants of the unperturbed problem in such a way that the secular terms\footnote{There is some ambiguity in identifying secular terms. Indeed, one can always add some regular terms to what one calls a secular term. This will result in a somewhat different renormalization flow equation. It may be important to make use of this freedom advantageously.} are exactly cancelled at the moment $\tau$. In the language of (\ref{nonlin}), one writes:
\beq
\begin{array}{l}
\dsty a=a(\tau,\epsilon)=a_R(\tau)+\epsilon a^{(1)}(a_R, b_R;\tau)+\epsilon^2 a^{(2)}(a_R, b_R;\tau)+\cdots,\vspace{2mm}\\ 
\dsty  b= b(\tau,\epsilon)= b_R(\tau)+\epsilon  b^{(1)}(a_R, b_R;\tau)+\epsilon^2  b^{(2)}(a_R, b_R;\tau)+\cdots,
\end{array}
\label{renormconst}
\eeq
where $a_R(\tau)$ and $ b_R(\tau)$ denote `renormalized integration constants' and we have omitted the mode number index. Note that the cancellation of secular terms at a given moment $\tau$ can always be arranged, simply because adjusting the initial conditions permits one to give the unperturbed trajectory absolutely any value at $\tau$. (\ref{renormconst}) have to be substituted to the na\"\i ve perturbative expansion and everything should be expressed through $a_R$ and $ b_R$.\vspace{2mm}

\noindent 2) One demands that the resulting perturbative expansion in terms of $a_R$ and $ b_R$ should be independent of $\tau$. Note that, once we introduce an $\epsilon$-dependence in the unperturbed solution in (\ref{renormconst}), we are no longer dealing with a single solution to the underlying problem, but with a family of asymptotic expansions. Demanding that the entire expansion is independent of $\tau$ simply amounts to forcing this family of asymptotic expansion to represent a single solution to the underlying problem (merely expanded in different ways), which is what we ultimately want to construct, rather than a family of solutions. Requiring the $\tau$-derivative of the expansion to vanish generates a first order differential equation for the renormalized integration constants,  $a_R(\tau)$ and $ b_R(\tau)$ for (\ref{renormconst}). This equation defines their renormalization flow.\vspace{2mm}

\noindent 3) After solving the renormalization flow equation obtained in 2), one substitutes the result in the expansion in terms of $a_R(\tau)$ and $ b_R(\tau)$, and finally sets $\tau$ to $t$ (this corresponds to working with a running coupling in perturbative quantum field theory). The result is free from secular terms by construction.\vspace{2mm}

In application to secular terms of the form (\ref{ressec}) at first non-trivial order  it is easy to see how the renormalization group method works. One may write the expansion as
\beq
a_i\cos(\om_it+ b_i)+\cdots+ \eps^{n}{\cal A}^{(n)}_i(a)\,t\sin(\om_i t+{\textstyle\sum_k} m_k b_k)+\cdots
\label{secmi}
\eeq
where we focus on the contribution of just one such term and the dots represent other terms, and $m_k$ is a certain set of integers. One first represents the secular term as
\beq
\eps^{n}{\cal A}^{(n)}_i(a)\,t\cos({\textstyle\sum_k} m_k b_k- b_i)\sin(\om_i t+ b_i)+\eps^{n}{\cal A}^{(n)}_i(a)\,t\sin({\textstyle\sum_k} m_k b_k- b_i)\cos(\om_i t+ b_i).
\eeq
Then one can absorb the secular term at moment $\tau$ by introducing
\beq
\begin{array}{l}
\dsty a_i=a_{R,i}-\eps^{n}{\cal A}^{(n)}_i(a_R)\,\tau\sin({\textstyle\sum_k} m_k b_{R,k}- b_{R,i}),\vspace{2mm}\\ 
\dsty  b_i= b_{R,i}+\frac{\eps^{n}}{a_i}{\cal A}^{(n)}_i(a_R)\,\tau\cos({\textstyle\sum_k} m_k b_{R,k}- b_{R,i}).
\end{array}
\label{barerenorm}
\eeq
As a result of re-expressing the expansion (\ref{secmi}) in terms of $a_R$ and $ b_R$, one gets 
\beq
\begin{array}{l}
\dsty a_{R,i}\cos(\om_it+ b_{R,i})+\cdots+\eps^{n}{\cal A}^{(n)}_i(a_R)\,(t-\tau)\cos({\textstyle\sum_k} m_k b_{R,k}- b_{R,i})\sin(\om_i t+ b_{R,i})\vspace{2mm}\\ 
\dsty \hspace{2cm}+\eps^{n}{\cal A}^{(n)}_i(a_R)\,(t-\tau)\sin({\textstyle\sum_k} m_k b_{R,k}- b_{R,i})\cos(\om_i t+ b_{R,i})+\cdots
\end{array}
\eeq
Equating to zero the $\tau$-derivative of this expression, on obtains the following renormalization flow equations:
\beq
\begin{array}{r}
\dsty \frac{da_{R,i}}{d\tau}=-\eps^{n}{\cal A}^{(n)}_i(a_R)\,\sin( b_{R,i}-{\textstyle\sum_k} m_k b_{R,k})+\cdots,\hspace{1mm}\vspace{2mm}\\ 
a_{R,i}\dsty \frac{d b_{R,i}}{d\tau}=-\eps^{n}{\cal A}^{(n)}_i(a_R)\,\cos( b_{R,i}-{\textstyle\sum_k} m_k b_{R,k})+\cdots,
\end{array}
\label{renorm}
\eeq
where the dots represent contributions from other secular terms. Note that the same equations would have resulted from formally differentiating (\ref{barerenorm}) with respect to $\tau$. In this way, one formally bypasses some of the steps in our above description of the method following \cite{Chen:1995ena}. (This is analogous to deriving the running of the renormalized coupling by differentiating the bare coupling with respect to the renormalization scale.)

For $N$ degrees of freedom, (\ref{renorm}) are $2N$ first order differential equations, a system of the same type\footnote{Note, however, that if a particular mode does not enter any frequency resonance relations $\sum_i m_i\om_i=0$, then this particular mode effectively decouples in (\ref{renorm}). The rank of the system is thereby reduced. The decoupling happens in the following way: for a completely non-resonant mode, the only secular terms allowed are (\ref{nonressecterm}). Such terms, according to (\ref{renorm}), induce an amplitude-dependent phase drift, but no amplitude drift. Similarly, the way a non-resonant mode enters the secular terms for the other modes is only through its (constant) amplitude, but not through its phase. Therefore, one can solve the equations for the entire set of resonant modes first, and then the result will simply contribute a slow phase drift to non-resonant modes. The Poincar\'e-Lindstedt method relies on an extreme version of this picture, when all the modes are non-resonant. Our main interest in this article is in non-linear AdS perturbations, a fully resonant system where none of such simplifications occur.} as our starting point (\ref{linsol}). Of course, (\ref{renorm}) contains less information than (\ref{linsol}) as it is entirely derived from a truncated perturbative expansion, whereas (\ref{linsol}) is exact. Nonetheless, one should generally not expect miraculous analytic solutions coming out of perturbative resummation under general circumstances. Equations (\ref{renorm}) are advantageous in that they explicitly describe very slow long-time energy flow between the different modes. They can thus be useful for analytic considerations of qualitative issues.

Having given a practical statement of the renormalization method in application to secular terms, we would like to zoom out for a moment and re-examine why and how the method works. When constructing na\"\i ve perturbation theory, one encounters growing terms at higher orders, which invalidate the perturbative expansion at large times. Nonetheless, one expects that the perturbative expansion is accurate for evolution over short times, before secular terms develop significant values. Logically, there is an obvious strategy here: to take the initial data at $t$, evolve them to $t+\Delta t$ using the na\"\i ve perturbation theory, build a new unperturbed solution starting from the values at $t+\Delta t$, develop a na\"\i ve perturbative expansion using this new solution, use it to evolve from $t+2\Delta t$, and so on, until one reaches the desired final time, which can be large. In this approach, one never uses the na\"\i ve perturbation theory outside its range of validity, and secular terms do not arise.

There is, in fact, more freedom than we have displayed in the construction from the previous passage. Indeed, when reaching $t+\Delta t$, we did not have to take the exact value of our variables as the initial value for the unperturbed solution on the interval $[t+\Delta t,t+2\Delta t]$. Rather, we could distribute this value in a convenient way between the initial value for the new unperturbed solution and the initial value for the perturbation. For example, we could only absorb the would-be secular terms into the new unperturbed solution on the interval $[t+\Delta t,t+2\Delta t]$, continuing the regular perturbative terms on the interval $[t,t+\Delta t]$ to the perturbation on the interval $[t+\Delta t,t+2\Delta t]$. We could also take $\Delta t$ to zero and implement these rearrangements in the perturbation series continuously on-the-go. This is precisely the picture underlying the renormalization group method. (Note that a slow running of the integration constants of the unperturbed solution emerges automatically in this perspective.)

The situation is directly analogous to what one encounters while dealing with ultraviolet divergences in perturbative field theory, with dependence on energy logarithm replacing the dependence on time. The na\"\i ve perturbative expansion for a quantity involving momenta of order $k$ regularized with a cut-off scale $\Lambda$, involves terms of the type $\ln(\Lambda^2/k^2)$. These terms become huge when the cut-off is sent to infinity and invalidate the perturbation theory (despite being suppressed by positive powers of the coupling). They are thus analogous to secular terms. One can formally introduce renormalized couplings at momentum scale $\mu$ related to the bare couplings by ill-behaved asymptotic series involving coefficients of the sort $\ln(\Lambda^2/\mu^2)$, so that the giant logarithms of the original perturbative expansion are replaced by  $\ln(\mu^2/k^2)$ when physical quantities are re-expressed through renormalized couplings. The new perturbation series is well-behaved for $k^2$ close to $\mu^2$, but ill-behaved elsewhere, when $\ln(\mu^2/k^2)$ (which is roughly analogous to $\Delta t$ of the previous passages) becomes large. Finally, one can demand that the whole construction should be independent of $\mu$, derive the corresponding renormalization group equation for the renormalized coupling dependence on $\mu$, solve them, and thereafer do all expansions at $\mu^2=k^2$ using the value of the renormalized coupling at $k^2$, thereby eliminating the (`secular') logarithms.

In fact, the renormalization method in application to secular term resummation operates in a mathematically much better defined setting than what quantum field theory may offer. One may hope to develop more transparent and tightly controlled derivations. We believe that starting from the time-stepping procedure described above, one should be able to derive the prescriptions of  \cite{Chen:1995ena} in a way that never involves ill-behaved asymptotic expansions with large coefficients at the intermediate steps. We shall nonetheless not pursue this program here.


\section{AdS (in)stability}\label{sec:AdS}

Having reviewed some possible strategies to deal with the problem of secular terms in na\"\i ve perturbation theory, we shall now turn to the issue of non-linear stability of the AdS space-time, in which such secular terms play a central role.

We shall work with the renormalization-based resummation of \cite{Chen:1995ena} at lowest non-trivial order. In principle, this approach is identical to the multiscale treatment of the same problem (at the same order) in \cite{Balasubramanian:2014cja}. Our perspective is quite different from \cite{Balasubramanian:2014cja}, however. There, the main focus was on numerical studies of the `energy flow' equations of the type (\ref{renorm}), truncated to a finite set of low-lying modes. Our main goal is to develop a neat analytic representation of these equations for all modes, with a view of future analytic studies of qualitative properties of this system. The main practical result we shall present here is the vanishing of an inifinite number of terms of a particular type in the flow equations, which are allowed on general grounds by the spectrum of frequencies of linear AdS perturbations.

\subsection{Setup of the system}

The equations of motion that we consider are Einstein's equations with a negative cosmological constant which are minimally coupled to a scalar field:
\begin{equation}
G_{\mu\nu}-\frac{d(d-1)}{2L^{2}}g_{\mu\nu}=8\pi G\left(\partial_{\mu}\phi\partial_{\nu}\phi-\frac{1}{2}g_{\mu\nu}(\partial\phi)^{2}\right)
\qquad\text{and}\qquad
\Box\phi=0.
\end{equation}
Following the conventions of \cite{Maliborski:2013jca}, we will parameterize the geometry by two functions $A(x,t)$ and $\delta(x,t)$ as
\begin{equation}
ds^{2}=\frac{L^{2}}{\cos^{2}x}\left(\frac{dx^{2}}{A}-Ae^{-2\delta}dt^{2}+\sin^{2}x\,d\Omega_{d-1}^{2}\right).
\end{equation}
The coordinates take values in $t\in]-\infty,\infty[$ and $x\in[0,\pi/2[$. The scalar field is also considered to be isotropic, $\phi=\phi(x,t)$. We introduce the notation $\Phi\equiv\phi'$ and $\Pi\equiv A^{-1}e^{\delta}\dot{\phi}$ (where overdots and primes denote derivatives with respect to $t$ and $x$, respectively) together with the convention $8\pi G=d-1$. Furthermore, it is convenient to define
\begin{equation}
\mu(x)\equiv(\tan x)^{d-1}
\qquad\text{and}\qquad
\nu(x)\equiv\frac{(d-1)}{\mu'(x)}=\frac{\sin x\cos x}{(\tan x)^{d-1}}.
\label{munu}
\end{equation}
The equations of motion then reduce to
\begin{subequations}\label{eq:eom}
\begin{align}
\dot{\Phi}&=\left(Ae^{-\delta}\Pi\right)',
&\dot{\Pi}&=\frac{1}{\mu}\left(\mu Ae^{-\delta}\Phi\right)', \\
A'&=\frac{\nu'}{\nu}\left(A-1\right)-\mu\nu\left(\Phi^{2}+\Pi^{2}\right)A,
&\delta'&=-\mu\nu\left(\Phi^{2}+\Pi^{2}\right),
\end{align}
\begin{equation}
\dot{A}=-2\mu\nu A^{2}e^{-\delta}\Phi\Pi.
\end{equation}
\end{subequations}
A static solution of these equations is the AdS-Schwarzschild black hole $A(x,t)=1-M\nu(x)$, $\delta(x,t)=0$ and $\phi(x,t)=0$. The unperturbed AdS space itself corresponds to $A=1$, $\de=\phi=0$.

\subsection{Weakly non-linear perturbation theory}

We will look for an approximate solution of the equations of motion (\ref{eq:eom}) with initial conditions $\phi(0,x)=\epsilon f(x)$ and $\dot{\phi}(0,x)=\epsilon g(x)$. Therefore, we expand the unknown functions in the amplitude of the initial conditions:
\begin{equation}
\phi(x,t)=\sum_{k=0}^{\infty}\epsilon^{2k+1}\phi_{2k+1}(x,t),
\,\,\,\,\,
A(x,t)=1+\sum_{k=1}^{\infty}\epsilon^{2k}A_{2k}(x,t),
\,\,\,\,\,
\delta(x,t)=\sum_{k=1}^{\infty}\epsilon^{2k}\delta_{2k}(x,t).
\end{equation}
At first order in the $\epsilon$-expansion, the equations of motion (\ref{eq:eom}) are linearized and result in the homogeneous partial differential equation
\begin{equation}
\ddot{\phi}_{1}+\hat{L}\phi_{1}=0
\qquad\text{with}\qquad
\hat{L}\equiv-\frac{1}{\mu(x)}\partial_{x}\left(\mu(x)\partial_{x}\right).
\end{equation}
The operator $\hat{L}$ is self-adjoint with respect to the inner product
\beq
 \langle\psi,\chi\rangle\equiv\int_{0}^{\pi/2}\bar{\psi}(x)\chi(x)\mu(x)\text{d}x.
\eeq 
The eigenvalues and eigenfunctions for $\hat{L}$ are $\omega_{j}^{2}=(d+2j)^{2}$ (with indices $j=0,1,2,...$) and
\begin{equation}
e_{j}(x)=k_{j}(\cos x)^{d}P_{j}^{\left(\frac{d}{2}-1,\frac{d}{2}\right)}\left(\cos(2x)\right)
\qquad\text{with}\qquad
k_{j}=\frac{2\sqrt{j!(j+d-1)!}}{\Gamma\left(j+\frac{d}{2}\right)}.
\label{ej}
\end{equation}
The function $P_{n}^{(a,b)}(x)$ is a Jacobi polynomial\footnote{Jacobi polynomials $P_{n}^{(a,b)}(x)$ are a system of orthogonal polynomials with respect to the measure $(1-x)^a(1+x)^b$ on the interval $(-1,1)$. A good summary of their properties with derivations is given in \cite{shentangwang}.} of order $n$. These eigenfunctions are defined such that $\hat{L}e_{j}=\omega_{j}^{2}e_{j}$ and $\langle e_{i},e_{j}\rangle=\delta_{ij}$. 
Note that all the mode frequencies $\om_j$ are integer and the spectrum is {\it fully resonant}, suggesting a large number of secular terms in non-linear perturbation theory.

We expand the unknown functions in the basis (\ref{ej}):
\begin{equation}
\phi_{2k+1}(x,t)=\sum_{j=0}^{\infty}c^{(2k+1)}_{j}(t)e_{j}(x)
\qquad\text{with}\qquad
c^{(2k+1)}_{j}(t)=\langle\phi_{2k+1}(x,t),e_{j}(x)\rangle.
\end{equation}
The solution of the linearized equation for $\phi_{1}$ is then given by
\begin{equation}
\phi_{1}(x,t)=\sum_{k=0}^{\infty}c_{j}(t)e_{j}(x). \label{eq:phi1}
\end{equation}
The coefficients $c_{j}\equiv c^{(1)}_{j}$ satisfy $\ddot{c}_{j}+\omega_{j}^{2}c_{j}=0$ and are thus given by 
\beq
c_{j}(t)=a_{j}\cos(\omega_{j}t+b_{j}),
\eeq
where the amplitudes $a_{j}$ and phases $b_{j}$ are determined by the initial profiles $f(x)$ and $g(x)$. The backreaction on the metric appears at second order. It is determined by the equations
\begin{subequations}
\begin{align}
A'_{2}&=\frac{\nu'}{\nu}A_{2}-\mu\nu\left((\dot{\phi}_{1})^{2}+(\phi'_{1})^{2}\right), \label{eq:A2prime} \\
\delta'_{2}&=-\mu\nu\left((\dot{\phi}_{1})^{2}+(\phi'_{1})^{2}\right), \label{eq:d2prime} \\
\dot{A}_{2}&=-2\mu\nu\dot{\phi}_{1}\phi'_{1}. \label{eq:A2dot}
\end{align}
\end{subequations}
These equations can be directly integrated to give
\begin{align}
A_{2}(x,t)&=-\nu(x)\int_{0}^{x}\left(\dot{\phi}_{1}(y,t)^{2}+\phi'_{1}(y,t)^{2}\right)\mu(y)\text{d}y, \label{eq:A2} \\
\delta_{2}(x,t)&=-\int_{0}^{x}\left(\dot{\phi}_{1}(y,t)^{2}+\phi'_{1}(y,t)^{2}\right)\mu(y)\nu(y)\text{d}y. \label{eq:d2}
\end{align}
At third order in the $\epsilon$-expansion, the equations of motion (\ref{eq:eom}) lead to the inhomogeneous equation
\begin{equation}\label{eq:S}
\ddot{\phi}_{3}+L\phi_{3}=S\equiv2\left(A_{2}-\delta_{2}\right)\ddot{\phi}_{1}+\left(\dot{A}_{2}-\dot{\delta}_{2}\right)\dot{\phi}_{1}+\left(A'_{2}-\delta'_{2}\right)\phi'_{1}.
\end{equation}
We can project this equation on to the eigenbasis $\{e_{j}\}$, such that
\begin{equation}\label{eq:modec3}
\ddot{c}^{(3)}_{j}+\omega_{j}^{2}c^{(3)}_{j}=S_{j}
\qquad\text{with}\qquad
S_{j}=\langle S,e_{j}\rangle.
\end{equation}
After a tedious but straightforward calculation (more details are explained in appendix \ref{sec:CalSl}), one finds the source term
\begin{align}\label{eq:Sl}
S_{l}&=\langle\left(A'_{2}-\delta'_{2}\right)\phi'_{1},e_{l}\rangle+2\langle A_{2}\ddot{\phi}_{1},e_{l}\rangle+\langle\dot{A}_{2}\dot{\phi}_{1},e_{l}\rangle-2\langle\delta_{2}\ddot{\phi}_{1},e_{l}\rangle-\langle\dot{\delta}_{2}\dot{\phi}_{1},e_{l}\rangle \nonumber \\
&=\frac{1}{2}\sum_{i=0}^{\infty}\sum_{\scriptsize{\begin{matrix}j=0\\j\neq i\end{matrix}}}^{\infty}\sum_{k=0}^{\infty}a_{i}a_{j}a_{k}\omega_{j}(H_{ijkl}-2X_{ijkl}\omega_{k}^{2})\left[\frac{1}{\omega_{i}-\omega_{j}}(\cos(\theta_{i}-\theta_{j}-\theta_{k})+\cos(\theta_{i}-\theta_{j}+\theta_{k}))\right. \nonumber \\
&\qquad\qquad\qquad\qquad\qquad\qquad\qquad\qquad\qquad\left.-\frac{1}{\omega_{j}+\omega_{i}}(\cos(\theta_{i}+\theta_{j}-\theta_{k})+\cos(\theta_{i}+\theta_{j}+\theta_{k}))\right] \nonumber \\
&-\frac{1}{4}\sum_{i=0}^{\infty}\sum_{k=0}^{\infty}a_{k}a_{i}^{2}(H_{iikl}-2\omega_{k}^{2}X_{iikl})(\cos(2\theta_{i}-\theta_{k})+\cos(2\theta_{i}+\theta_{k})) \nonumber \\
&-\frac{1}{2}\sum_{i=0}^{\infty}\sum_{k=0}^{\infty}a_{k}a_{i}^{2}(H_{iikl}+2\omega_{i}^{2}M_{kli}-2\omega_{k}^{2}X_{iikl}-4\omega_{k}^{2}\omega_{i}^{2}W_{kli})\cos(\theta_{k}) \nonumber \\
&-\frac{1}{2}\sum_{i=0}^{\infty}\sum_{j=0}^{\infty}\sum_{k=0}^{\infty}X_{ijkl}a_{i}a_{j}a_{k}\omega_{j}\omega_{k}\left[\cos(\theta_{k}-\theta_{j}-\theta_{i})+\cos(\theta_{k}-\theta_{j}+\theta_{i})\right. \nonumber \\
&\qquad\qquad\qquad\qquad\qquad\qquad\qquad\left.-\cos(\theta_{k}+\theta_{j}-\theta_{i})-\cos(\theta_{k}+\theta_{j}+\theta_{i})\right] \nonumber \\
&+\frac{1}{4}\sum_{\scriptsize{\begin{matrix}k=0\\k\neq l\end{matrix}}}^{\infty}\sum_{i=0}^{\infty}\sum_{j=0}^{\infty}\frac{a_{i}a_{j}a_{k}\omega_{k}}{(\omega_{l}^{2}-\omega_{k}^{2})}\left\{Z^{+}_{ijkl}(2\omega_{k}+\omega_{j}-\omega_{i})\cos(\theta_{i}-\theta_{j}-\theta_{k})\right. \nonumber \\
&\qquad\qquad\qquad\qquad\qquad\qquad\left.+Z^{+}_{ijkl}(2\omega_{k}-\omega_{j}+\omega_{i})\cos(\theta_{i}-\theta_{j}+\theta_{k})\right. \nonumber \\
&\qquad\qquad\qquad\qquad\qquad\qquad\left.+Z^{-}_{ijkl}(\omega_{i}+\omega_{j}-2\omega_{k})\cos(\theta_{i}+\theta_{j}-\theta_{k})\right. \nonumber \\
&\qquad\qquad\qquad\qquad\qquad\qquad\left.-Z^{-}_{ijkl}(2\omega_{k}+\omega_{j}+\omega_{i})\cos(\theta_{i}+\theta_{j}+\theta_{k})\right\} \nonumber \\
&-\frac{1}{4}\sum_{i=0}^{\infty}\sum_{j=0}^{\infty}a_{i}a_{j}a_{l}\omega_{l}\left\{[\omega_{i}\omega_{j}P_{ijl}+B_{ijl}](2\omega_{l}+\omega_{j}-\omega_{i})\cos(\theta_{i}-\theta_{j}-\theta_{l})\right. \nonumber \\
&\qquad\qquad\qquad\qquad\left.+[\omega_{i}\omega_{j}P_{ijl}+B_{ijl}](2\omega_{l}-\omega_{j}-\omega_{i})\cos(\theta_{i}-\theta_{j}+\theta_{l})\right. \nonumber \\
&\qquad\qquad\qquad\qquad\left.+[\omega_{i}\omega_{j}P_{ijl}-B_{ijl}](\omega_{j}+\omega_{i}-2\omega_{l})\cos(\theta_{i}+\theta_{j}-\theta_{l})\right. \nonumber \\
&\qquad\qquad\qquad\qquad\left.-[\omega_{i}\omega_{j}P_{ijl}-B_{ijl}](2\omega_{l}+\omega_{j}+\omega_{i})\cos(\theta_{i}+\theta_{j}+\theta_{l})\right\},
\end{align}
where we used the shorthand notation $\theta_{i}(t)=\omega_{i}t+b_{i}$. The coefficients that appear in these expressions are certain integrals of products of eigenfunctions:
\begin{subequations}\label{eq:coefs}
\begin{align}
H_{ijkl}&=\int_{0}^{\frac{\pi}{2}}\text{d}x\,e'_{i}(x)e_{j}(x)e'_{k}(x)e_{l}(x)(\mu(x))^{2}\nu'(x) \label{Hijkl},\\
X_{ijkl}&=\int_{0}^{\frac{\pi}{2}}\text{d}x\,e'_{i}(x)e_{j}(x)e_{k}(x)e_{l}(x)(\mu(x))^{2}\nu(x), \\
Y_{ijkl}&=\int_{0}^{\frac{\pi}{2}}\text{d}x\,e'_{i}(x)e_{j}(x)e'_{k}(x)e'_{l}(x)(\mu(x))^{2}\nu(x), \\
Z^{\pm}_{ijkl}&=\omega_{i}\omega_{j}(X_{klij}-X_{lkij})\pm(Y_{klij}-Y_{lkij}),\label{Zpm}\\
M_{ijk}&=\int_{0}^{\frac{\pi}{2}}\text{d}x\,e'_{i}(x)e_{j}(x)\mu(x)\nu'(x)\int_{0}^{x}\text{d}y(e_{k}(y))^{2}\mu(y), \\
W_{ijk}&=\int_{0}^{\frac{\pi}{2}}\text{d}x\,e_{i}(x)e_{j}(x)\mu(x)\nu(x)\int_{0}^{x}\text{d}y(e_{k}(y))^{2}\mu(y), \\
P_{ijk}&=\int_{0}^{\frac{\pi}{2}}\text{d}x\,e_{i}(x)e_{j}(x)\mu(x)\nu(x)\left(1-\int_{0}^{x}\text{d}y(e_{k}(y))^{2}\mu(y)\right), \\
B_{ijk}&=\int_{0}^{\frac{\pi}{2}}\text{d}x\,e'_{i}(x)e'_{j}(x)\mu(x)\nu(x)\left(1-\int_{0}^{x}\text{d}y(e_{k}(y))^{2}\mu(y)\right).
\end{align}
\end{subequations}

\subsection{Vanishing secular terms}\label{vanish}

As already discussed above (\ref{secgen}), secular terms appear when the set of frequencies $\{\omega_{i},\omega_{j},\omega_{k}\}$ satisfies the resonance condition $\omega_{i}\pm\omega_{j}\pm\omega_{k}=\pm\omega_{l}$. In this case, a resonant term should (generally) arise in the source $S_{l}$ of the mode $c^{(3)}_{l}$. Equation (\ref{eq:modec3}) will then have a solution that involves a secular term,
\begin{equation}
\ddot{c}^{(3)}_{l}(t)+\omega_{l}^{2}c^{(3)}_{l}(t)={\cal A}\cos(\omega_{l}t+{\cal B})+(...)
\quad\Rightarrow\quad
c^{(3)}_{l}(t)=\frac{\cal A}{2\omega_{l}}t\sin(\omega_{l}t+{\cal B})+(...).
\label{sourcetosec}
\end{equation}
There are eight choices of the signs in $\omega_{i}\pm\omega_{j}\pm\omega_{k}=\pm\omega_{l}$.  First, one can have $\omega_{i}+\omega_{j}+\omega_{k}=\omega_{l}$. We shall call the corresponding terms `+++ terms'. $\omega_{i}+\omega_{j}+\omega_{k}=-\omega_{l}$ cannot be satisfied due to frequency positivity. Of the remaining six choices, three can be brought to the form $\omega_{i}+\omega_{j}-\omega_{k}=\omega_{l}$ by permuting $i$, $j$ and $k$. We shall call these `++\,- terms'. After that, the three remaining choices can be brought to the form $\omega_{i}-\omega_{j}-\omega_{k}=\omega_{l}$ by permuting $i$, $j$ and $k$. We shall call these `+\,-\,- terms'.

Our goal in this section is to prove that the +++ and +\,-\,- terms vanish due to properties of the AdS mode functions, despite being allowed by the frequency spectrum. We shall give explicit expressions for the ++\,- terms in section \ref{nonvanish}.

We shall first focus on the +++ terms in (\ref{eq:Sl}), for which $\omega_{i}+\omega_{j}+\omega_{k}=\omega_{l}$. One finds that
\begin{equation}
S_{l}=(...)+\underbrace{\sum_{i}\sum_{j}\sum_{k}}_{i+j+k+d=l}Q_{ijkl}a_{i}a_{j}a_{k}\cos(\theta_{i}+\theta_{j}+\theta_{k}),
\end{equation}
where $(...)$ represents the non-resonant terms as well as resonant terms of other types. The coefficients $Q_{ijkl}$ are given by
\begin{align}
Q_{ijkl}=-\frac{1}{12}H_{ijkl}&\frac{\omega_{j}(2\omega_{j}+\omega_{i}+\omega_{k})}{(\omega_{j}+\omega_{i})(\omega_{j}+\omega_{k})}-\frac{1}{12}H_{jkil}\frac{\omega_{k}(2\omega_{k}+\omega_{i}+\omega_{j})}{(\omega_{k}+\omega_{i})(\omega_{k}+\omega_{j})} \nonumber \\
&-\frac{1}{12}H_{kijl}\frac{\omega_{i}(2\omega_{i}+\omega_{j}+\omega_{k})}{(\omega_{i}+\omega_{j})(\omega_{i}+\omega_{k})}+\frac{1}{6}X_{ijkl}\,\omega_{j}\omega_{k}\left(1+\frac{\omega_{k}}{(\omega_{j}+\omega_{i})}+\frac{\omega_{j}}{(\omega_{k}+\omega_{i})}\right) \nonumber \\
+\frac{1}{6}X_{jkil}\,\omega_{i}\omega_{k}&\left(1+\frac{\omega_{k}}{(\omega_{i}+\omega_{j})}+\frac{\omega_{i}}{(\omega_{k}+\omega_{j})}\right)+\frac{1}{6}X_{kijl}\,\omega_{i}\omega_{j}\left(1+\frac{\omega_{i}}{(\omega_{j}+\omega_{k})}+\frac{\omega_{j}}{(\omega_{i}+\omega_{k})}\right) \nonumber \\
&-\frac{1}{12}Z^{-}_{ijkl}\frac{\omega_{k}}{(\omega_{i}+\omega_{j})}-\frac{1}{12}Z^{-}_{jkil}\frac{\omega_{i}}{(\omega_{j}+\omega_{k})}-\frac{1}{12}Z^{-}_{kijl}\frac{\omega_{j}}{(\omega_{i}+\omega_{k})}.
\label{ppp}
\end{align}
We will now show that all coefficients $Q_{ijkl}$ vanish whenever $i+j+k+d=l$, i.e., when the resonance condition is satisfied. This is a non-trivial statement (though it may well have a more straightforward and elegant proof based on the symmetries of AdS), since the structure of the linearized frequency spectrum allows such terms.

To analyze (\ref{ppp}) we employ the following transformations:\vspace{2mm}

\noindent {\bf 1)} We notice that
\begin{equation}
H_{ijkl}=\omega_{k}^{2}X_{ijkl}-Y_{ijkl}+\omega_{i}^{2}X_{klij}-Y_{klij}.
\label{Hid}
\end{equation}
This identity is proved by integration by parts removing the derivative from $\nu$ in (\ref{Hijkl}).
Antisymmetrizing (\ref{Hid}) with respect to $i$ and $j$, one gets:
\beq
Y_{ijkl}-Y_{jikl}=(\omega_i^2-\omega_j^2)X_{klij}+\omega_k^2(X_{ijkl}-X_{jikl})-(H_{ijkl}-H_{jikl}).
\label{Yint}
\eeq
This relation is used to eliminate $Y$ from $Z^+$ of (\ref{Zpm}), and hence from (\ref{ppp}).
\vspace{2mm}

\noindent {\bf 2)} After the above manipulation, (\ref{ppp}) only contains $H$ and $X$ of (\ref{eq:coefs}). $H$ and $X$ are integrals of products of mode functions $e_i$, $e_j$, $e_k$, $e_l$ and their derivatives. An important distinction among the different terms is whether $e_l$ (in the integrand of $H$ or $X$) is differentiated. ($e_l$ is the mode function of the mode receiving the secular term contribution.)  If it is, we remove the derivative from it using integration by parts:
\beq
\begin{array}{l}
\dsty H_{lijk}=-(H_{ikjl}+H_{kijl})+\om_j^2D_{ijkl}+4X_{jikl},\vspace{2mm}\\
\dsty X_{lijk}=-(X_{ijkl}+X_{jikl}+X_{kijl})-E_{ijkl},
\end{array}
\label{HXint}
\eeq
with
\beq
\begin{array}{l}
\dsty D_{ijkl}=\int_{0}^{\frac{\pi}{2}}\text{d}x\,e_ie_je_ke_l\mu^2\nu',\vspace{2mm}\\
\dsty E_{ijkl}=\int_{0}^{\frac{\pi}{2}}\text{d}x\,e_ie_je_ke_l(\mu^2\nu)'.
\end{array}
\eeq
(We have used the identity $(\mu\nu')'=-4\mu\nu$.) At this point,\footnote{One could have skipped directly from (\ref{ppp}) to (\ref{elqijk}), though the integrands would have involved more derivatives than what we get after having performed the integrations by parts and are less convenient to handle. For the case of +\,-\,- terms we consider below, the integrations by parts we have described are necessary to establish the analog of (\ref{elqijk}).} (\ref{ppp}) takes the form
\beq
Q_{ijkl}\sim \int_{0}^{\frac{\pi}{2}}\text{d}x\,e_l(x)q_{ijk}(x),
\label{elqijk}
\eeq
where $q_{ijk}(x)$ can only receive the following contributions: from $H$-terms, a product of $e_i$, $e_j$, $e_k$, two of which are differentiated, times $\mu^2\nu'$; from $X$-terms, a product of $e_i$, $e_j$, $e_k$, one of which is differentiated, times $\mu^2\nu$; from $D$-terms, $e_ie_je_ke_l\mu^2\nu'$; from $E$-terms, $e_ie_je_ke_l(\mu^2\nu)'$.\vspace{2mm}

\noindent {\bf 3)} From (\ref{elqijk}) and the expression for mode functions in terms of Jacobi polynomials (\ref{ej}), one can bring (\ref{ppp}) to the form\footnote{Each of the types of terms listed under (\ref{elqijk}) is individually of this form.}
\beq
Q_{ijkl}\sim \int_{-1}^{1}P_l^{\left(\frac{d}{2}-1,\frac{d}{2}\right)}(\xi) {\cal Q}_{ijk}(\xi) (1-\xi)^{\frac{d}2-1}(1+\xi)^{\frac{d}2}d\xi
\label{Qpoly}
\eeq
with $\xi=\cos 2x$. Here, ${\cal Q}_{ijk}(\xi)$ is a polynomial of degree $i+j+k+d+1=l+1$ made of $P_i^{\left(\frac{d}{2}-1,\frac{d}{2}\right)}$, $P_j^{\left(\frac{d}{2}-1,\frac{d}{2}\right)}$, $P_k^{\left(\frac{d}{2}-1,\frac{d}{2}\right)}$, their first derivatives and various trigonometric functions appearing in  (\ref{elqijk}), re-expressed through $\cos 2x$. Note that the integration measure appearing in (\ref{Qpoly}) is precisely the same as the one used for defining the Jacobi polynomials.\vspace{2mm}

\noindent {\bf 4)} Finally, to prove that (\ref{Qpoly}) vanishes, it suffices to show that the expansion of ${\cal Q}_{ijk}(\xi)$ in terms of Jacobi polynomials  $P_n^{\left(\frac{d}{2}-1,\frac{d}{2}\right)}$ does not contain $P_l^{\left(\frac{d}{2}-1,\frac{d}{2}\right)}$. Since ${\cal Q}_{ijk}(\xi)$ is a polynomial of degree $l+1$, whether that happens or not can be decided on the basis of considering the coefficients of its two highest powers. More specifically, the coefficients of the two highest powers in Jacobi polynomials (which we need only up to the overall normalization) can be extracted from known formulas (see, e.g., \cite{shentangwang}) as
\beq
P_n^{(d/2-1,d/2)}(\xi)\sim (1+\xi)^n-\frac{n(d+2n)}{d+2n-1}(1+\xi)^{n-1}+\cdots
\label{1x}
\eeq
If one uses this representation for $P_i^{\left(\frac{d}{2}-1,\frac{d}{2}\right)}$, $P_j^{\left(\frac{d}{2}-1,\frac{d}{2}\right)}$, $P_k^{\left(\frac{d}{2}-1,\frac{d}{2}\right)}$  to recover the coefficients of the two highest powers in ${\cal Q}_{ijk}$, one finds
\beq
{\cal Q}_{ijk}\sim (1+\xi)^{i+j+k+d+1}-\tfrac{(d+i+j+k+1)(3d+2i+2j+2k+2)}{3d+2i+2j+2k+1}(1+\xi)^{i+j+k+d}+\cdots,
\label{Qcomp}
\eeq
which exactly matches (\ref{1x}) with $n=l+1=i+j+k+d+1$. Therefore, if one subtracts from ${\cal Q}_{ijk}$ its projection on $P_{l+1}^{(d/2-1,d/2)}$, the remaining polynomial is of degree $l-1$ and cannot have a non-zero projection on $P_l^{(d/2-1,d/2)}$. Then, from (\ref{Qpoly}),
\beq
Q_{ijkl}=0.
\eeq

We have thus proved that all secular terms resulting from addition of three mode frequencies vanish for non-linear perturbation theory in the AdS background, even if the said combination of frequencies resonates with another perturbation mode. In relation to our proof sketched above, it remains only to comment on how one in practice computes the polynomial ${\cal Q}_{ijk}(\xi)$ in (\ref{Qpoly}), and thus establishes (\ref{Qcomp}).

The computation reconstructing ${\cal Q}_{ijk}(\xi)$ in (\ref{Qpoly}) is, in principle, a completely straightforward polynomial manipulation, but it is the forbiddingly large size of the polynomial expressions that makes the manipulations demanding. For the special case $i=j=k$, we have been able to do the entire computation by hand and derive (\ref{Qcomp}). However, for arbitrary $i$, $j$ and $k$ one has to either invent powerful analytic tricks, perform pages upon pages of completely mechanical polynomial manipulations, or resort to (fully analytic) computer algebra. For the purposes of this article, we have chosen the latter and employed FORM, a powerful script-based symbolic manipulation system particularly suited for working with long polynomial expressions \cite{FORM}. Our FORM script, essentially retracing the steps presented above in this section, is given in Appendix \ref{sec:form}. (\ref{Qcomp}) can be read off from the output of that script, thereby completing our proof.

We now turn to the +\,-\,- terms with $\omega_{i}-\omega_{j}-\omega_{k}=\omega_{l}$. These terms vanish in a way very similar to what we have just observed for the +++ terms. The corresponding contribution to the source $S_l$ is given by
\begin{equation}
S_{l}=(...)+\sum_{j}\sum_{k}U_{(j+k+l+d)jkl}a_{j+k+l+d}a_{j}a_{k}\cos(\theta_{j+k+l+d}-\theta_{j}-\theta_{k})
\end{equation}
with the following coefficients:
\begin{align}
U_{ijkl}&=\frac{1}{4}H_{ijkl}\frac{\omega_{j}(2\omega_{j}-\omega_{i}+\omega_{k})}{(\omega_{i}-\omega_{j})(\omega_{j}+\omega_{k})}+\frac{1}{4}H_{jkil}\frac{\omega_{k}(2\omega_{k}-\omega_{i}+\omega_{j})}{(\omega_{i}-\omega_{k})(\omega_{k}+\omega_{j})}+\frac{1}{4}H_{kijl}\frac{\omega_{i}(\omega_{j}+\omega_{k}-2\omega_{i})}{(\omega_{i}-\omega_{j})(\omega_{i}-\omega_{k})} \nonumber \\
&-\frac{1}{2}X_{ijkl}\,\omega_{j}\omega_{k}\left(\frac{\omega_{k}}{(\omega_{i}-\omega_{j})}+\frac{\omega_{j}}{(\omega_{i}-\omega_{k})}-1\right)+\frac{1}{2}X_{jkil}\,\omega_{i}\omega_{k}\left(\frac{\omega_{k}}{(\omega_{i}-\omega_{j})}+\frac{\omega_{i}}{(\omega_{k}+\omega_{j})}-1\right) \nonumber \\
&+\frac{1}{2}X_{kijl}\,\omega_{i}\omega_{j}\left(\frac{\omega_{i}}{(\omega_{j}+\omega_{k})}+\frac{\omega_{j}}{(\omega_{i}-\omega_{k})}-1\right) \nonumber \\
&-\frac{1}{4}Z^{+}_{ijkl}\frac{\omega_{k}}{(\omega_{i}-\omega_{j})}+\frac{1}{4}Z^{-}_{jkil}\frac{\omega_{i}}{(\omega_{j}+\omega_{k})}-\frac{1}{4}Z^{+}_{kijl}\frac{\omega_{j}}{(\omega_{i}-\omega_{k})}.
\label{ppm}
\end{align}
One can show that $U_{ijkl}=0$ whenever the resonance condition is satisfied, i.e., $i=j+k+l+d$, in other words, the +\,-\,- terms do not arise.

One can construct a proof that $U_{ijkl}=0$  essentially repeating the procedure we employed above for the +++ terms, except that the roles of $i$ and $l$ become interchanged. Performing appropriate integrations by parts using (\ref{Yint}-\ref{HXint}), one arrives at the following representation:
\beq
U_{ijkl}\sim \int_{-1}^{1}P_i^{\left(\frac{d}{2}-1,\frac{d}{2}\right)}(\xi)\, {\cal U}_{jkl}(\xi) (1-\xi)^{\frac{d}2-1}(1+\xi)^{\frac{d}2}d\xi,
\label{Upoly}
\eeq
where ${\cal U}_{jkl}(\xi)$ is a polynomial of degree $j+l+k+d+1=i+1$. Substituting explicit expressions for the mode functions, one finds that the coefficients of the two highest powers in ${\cal U}_{jkl}(\xi)$ are in the same ratio as in  $P_{i+1}^{\left(\frac{d}{2}-1,\frac{d}{2}\right)}$ (in practice, we have used a FORM script to perform this polynomial evaluation, and the output of the script is given at the end of Appendix~\ref{sec:form}). ${\cal U}_{jkl}(\xi)$ is then orthogonal to $P_i^{\left(\frac{d}{2}-1,\frac{d}{2}\right)}(\xi)$ and (\ref{Upoly}) vanishes, which completes our proof of the absense of the +\,-\,- secular terms.

It would be desirable to develop a more elegant and less computationally intensive proof of the vanishing of the +++ and +\,-\,- terms, and indeed understand the qualitative reason for these terms to vanish. In section \ref{sec:qp} we make some preliminary comments on the relation between the absence of these classes of secular terms and abundance of quasiperiodic solutions to the full non-linear system.

\subsection{Non-vanishing secular terms and renormalization flow}\label{nonvanish}

The non-vanishing (++\,-) resonant terms in the source $S_{l}$ arise from resonances that have the form $\omega_{i}+\omega_{j}-\omega_{k}=\omega_{l}$ and are given by
\begin{align}
S_{l}=(...)&+T_{l}a_{l}^{3}\cos(\theta_{l}+\theta_{l}-\theta_{l})+\sum_{i,(i\neq l)}R_{il}a_{i}^{2}a_{l}\cos(\theta_{i}+\theta_{l}-\theta_{i}) \nonumber \\
&+\underbrace{\sum_{i,(i\neq l)}\sum_{j,(j\neq l)}}_{l\leqslant i+j}S_{ij(i+j-l)l}a_{i}a_{j}a_{i+j-l}\cos(\theta_{i}+\theta_{j}-\theta_{i+j-l}),
\end{align}
where $(...)$ represents the non-resonant terms. The coefficients $S_{ijkl}$, $R_{il}$ and $T_{l}$ are given by
\begin{align}\label{SRT1}
S_{ijkl}=&-\frac{1}{4}H_{ijkl}\omega_{j}\left(\frac{1}{\omega_{j}+\omega_{i}}+\frac{1}{\omega_{j}-\omega_{k}}\right)-\frac{1}{4}H_{jkil}\omega_{k}\left(\frac{1}{\omega_{k}-\omega_{i}}+\frac{1}{\omega_{k}-\omega_{j}}\right) \nonumber \\
&-\frac{1}{4}H_{kijl}\omega_{i}\left(\frac{1}{\omega_{i}+\omega_{j}}+\frac{1}{\omega_{i}-\omega_{k}}\right)+\frac{1}{2}X_{kijl}\omega_{i}\omega_{j}\left(\frac{\omega_{j}}{\omega_{i}-\omega_{k}}+\frac{\omega_{i}}{\omega_{j}-\omega_{k}}+1\right) \nonumber \\
&+\frac{1}{2}X_{ijkl}\omega_{j}\omega_{k}\left(\frac{\omega_{k}}{\omega_{j}+\omega_{i}}+\frac{\omega_{j}}{\omega_{k}-\omega_{i}}-1\right)+\frac{1}{2}X_{jkil}\omega_{k}\omega_{i}\left(\frac{\omega_{k}}{\omega_{i}+\omega_{j}}+\frac{\omega_{i}}{\omega_{k}-\omega_{j}}-1\right) \nonumber \\
&+\frac{1}{4}\left(\frac{\omega_{k}}{\omega_{i}+\omega_{j}}\right)Z^{-}_{ijkl}+\frac{1}{4}\left(\frac{\omega_{i}}{\omega_{j}-\omega_{k}}\right)Z^{+}_{jkil}+\frac{1}{4}\left(\frac{\omega_{j}}{\omega_{i}-\omega_{k}}\right)Z^{+}_{kijl}, \\
R_{il}=&
\left(\frac{\omega_{i}^{2}}{\omega_{l}^{2}-\omega_{i}^{2}}\right)\left(H_{liil}-2\omega_{i}^{2}X_{liil}\right)-\left(\frac{\omega_{l}^{2}}{\omega_{l}^{2}-\omega_{i}^{2}}\right)\left(H_{ilil}-2\omega_{i}^{2}X_{ilil}\right)-\omega_{i}^{2}X_{liil} \nonumber \\
&-\frac{1}{2}\left(H_{iill}+2\omega_{i}^{2}M_{lli}\right)+\omega_{l}^{2}\left(X_{iill}+2\omega_{i}^{2}W_{lli}\right)-\omega_{l}^{2}\left(\omega_{i}^{2}P_{iil}+B_{iil}\right) \nonumber \\
&+\left(\frac{\omega_{i}^{2}}{\omega_{l}^{2}-\omega_{i}^{2}}\right)\left(Y_{illi}-Y_{lili}+\omega_{l}^{2}(X_{illi}-X_{lili})\right), \\
\label{SRT2}T_{l}=&-\frac{3}{4}H_{llll}+\omega_{l}^{2}X_{llll}-\omega_{l}^{2}M_{lll}-\omega_{l}^{2}B_{lll}+2\omega_{l}^{4}W_{lll}-\omega_{l}^{4}P_{lll}.
\end{align}
As per (\ref{sourcetosec}), to convert these source terms to secular terms in the solution for $c^{(3)}_{l}$, one simply needs to replace all cosines by sines, and multiply by $t/(2\om_l)$.
From such an expression for the secular terms, retracing the steps between (\ref{secmi}) and (\ref{renorm}), one obtains the following renormalization flow equations for non-linear perturbation theory in the AdS background at first non-trivial order:
\begin{align}
\frac{2\om_l}{\epsilon^2}\frac{dA_{l}}{dt}=&-\underbrace{\sum_{i,(i\neq l)}\sum_{j,(j\neq l)}}_{l\leqslant i+j}S_{ij(i+j-l)l}A_{i}A_{j}A_{i+j-l}\sin(B_l+B_{i+j-l}-B_{i}-B_{j}),\label{renAl}\\
\frac{2\om_lA_l}{\epsilon^2}\frac{dB_{l}}{dt}=&-T_{l}A_{l}^{3}-\sum_{i,(i\neq l)}R_{il}A_{i}^{2}A_{l}\nonumber \\
&-\underbrace{\sum_{i,(i\neq l)}\sum_{j,(j\neq l)}}_{l\leqslant i+j}S_{ij(i+j-l)l}A_{i}A_{j}A_{i+j-l}\cos(B_l+B_{i+j-l}-B_{i}-B_{j}),\label{renBl}
\end{align}
where $A_l$ and $B_l$ are the (slowly) running renormalized amplitudes and phases, and the numerical coefficients $T$, $R$ and $S$ can be read off (\ref{munu}), (\ref{ej}), (\ref{eq:coefs}), (\ref{SRT1}-\ref{SRT2}).

\subsection{Renormalization flow and quasi-periodic solutions}\label{sec:qp}

Numerical investigations of \cite{Bizon:2011gg,Dias:2011ss,Dias:2012tq,Maliborski:2013jca,Buchel:2013uba,Abajo-Arrastia:2014fma,Maliborski:2014rma,Balasubramanian:2014cja} have revealed a complex interplay between stability and instability depending on the shape of the initial AdS perturbation. We feel that this feature finds a reflection in the weakly non-linear perturbation theory, since, despite the fact that the frequency spectrum is fully resonant (and thus, for example, no orbits at all are protected from instability by the KAM theorem), only a subset of possible secular terms (an correspondingly, energy transfer channels in the renormalization flow equations) actually appear.

To make this more precise, we revisit the perturbative analysis of quasiperiodic solutions in \cite{Balasubramanian:2014cja}. In that article, `Two-Time Framework' equations identical to our (\ref{renAl}-\ref{renBl}) were derived. The coefficients were not given analytic expressions, but rather evaluated explicitly using a computer for a system truncated to low-lying modes. The vanishing of the +++ and +\,-\,- secular term, for which we have given an analytic proof in section \ref{vanish}, was of course observed (for a particular set of low-lying modes) in the results of those direct evaluations. The authors then asked whether their `Two-Time Framework' equation predict solutions that remain quasi-periodic for times of order $1/\epsilon^2$ (which is the validity range of the resummed perturbation theory).

We can ask the same quasi-periodicity question in the context of our system (\ref{renAl}-\ref{renBl}), and also contemplate how the presence of more general terms in (\ref{renAl}-\ref{renBl}) would have affected the abundance of quasi-periodic solutions. More general terms in (\ref{renAl}-\ref{renBl}) could be there given the frequency spectrum of AdS perturbation but are in fact absent due to the vanishing of some classes of secular terms specific to the AdS background, which we have analyzed in section \ref{vanish}.

Quasiperiodicity in the language of (\ref{renAl}-\ref{renBl}) simply means that the renormalized amplitudes $A_l$ are constant. In that case, there is no significant energy transfer between the modes (small energy oscillations are produced by non-secular terms in perturbation theory), and the only significant effect of non-linearities on the evolution is the linear drift of the renormalized phases $B_l$ due to (\ref{renBl}) which is nothing but a Poincar\'e-Lindstedt frequency shift. (This picture of quasi-periodic motion in a non-linear non-integrable system is familiar from the KAM theory, though the fully resonant frequency spectrum we are dealing with is exactly the opposite of the KAM theory asumptions.)

$A_l$ in  (\ref{renAl}) will vanish if
\beq
B_l+B_{i+j-l}-B_{i}-B_{j}=0
\label{phaseeq}
\eeq
for all $l$, $i$, $j$. As observed already in \cite{Balasubramanian:2014cja}, this is solved by
\beq
B_j=B_0+j(B_1-B_0),
\label{phasesol}
\eeq
where $B_0$ and $B_1$ can be arbitrary. One then substitutes this relation into (\ref{renBl}) and obtains a system of algebraic equations for $B_0$, $B_1$ and $A_l$. As pointed out in \cite{Balasubramanian:2014cja}, if one truncates this system to a finite subset of low-lying modes up to $j=j_{max}$, one obtains $j_{max}+1$ equations for $j_{max}+3$ equations, giving a 2-parameter family of solutions, which actually becomes a 1-parameter family of essentially different quasi-periodic solutions after the obvious scaling symmetry $A_l(t)\to \xi A_l(t/\xi^2)$, $B_l(t)\to B_l(t/\xi^2)$ present in (\ref{renAl}-\ref{renBl}) is taken into account. (Removing the mode cut-off is subtle and we shall not attempt to do it carefully at present.)

What we would like to emphasize in the context of our study is that the situation would have changed if more general terms (that vanish specifically for the AdS case) were present on the right-hand side of (\ref{renAl}). Such terms would have different dependences on phases. For example, the +++ terms of section \ref{vanish} would have produced $\sin(B_l-B_i-B_j-B_{l-d-i-j})$ and the +\,-\,- terms would have produced $\sin(B_l+B_i+B_j-B_{l+i+j+d})$. Demanding these terms to vanish would produce more equations, in addition to (\ref{phaseeq}), which can in general only be solved by
\beq
B_j=\frac{\om_j}{\om_0}B_0=\left(1+\frac{2j}d\right)B_0.
\eeq
This equation contains only one free parameter, $B_0$, whereas (\ref{phasesol}) contains two, $B_0$ and $B_1$. Consequently, based on this simple counting we observe that the number of free parameters labelling different quasiperiodic solutions diminishes by one when general secular terms are present, compared to the AdS case, where the +++ and +\,-\,- term vanish. This observation gives some non-perturbative meaning to the restrictions on the type of secular terms appearing in the AdS perturbation theory. It would be good to make these ideas more precise.


\section{Acknowledgments}

We would like to thank Piotr Bizo\'n, Alex Buchel, Luis Lehner and Andrzej Rostworowski for useful discussions. The work of B.C.\ and J.V.\ has been supported by the Belgian Federal Science Policy Office through the Interuniversity Attraction Pole P7/37, by FWO-Vlaanderen through project G020714N, and by the Vrije Universiteit Brussel through the Strategic Research Program ``High-Energy Physics.'' The research of O.E.\ has been supported by Ratchadaphisek Sompote Endowment Fund. J.V.\ is supported by a PhD Fellowship of the Research Foundation Flanders (FWO).

\appendix

\section{Calculation of $S_{l}$}\label{sec:CalSl}

In this section, we give details on the computation of $S_{l}=\langle S,e_{l}\rangle$. Before we start, we list some useful identities and definitions. From the equation $\ddot{c}_{j}+\omega_{j}^{2}c_{j}=0$ for the modes $c_{j}(t)$ it follows that $\frac{d}{dt}(\omega_{i}^{2}c_{i}^{2}+\dot{c}_{i}^{2})=0$. Therefore, we can define the constants
\begin{equation}
C_{i}=\omega_{i}^{2}c_{i}^{2}+\dot{c}_{i}^{2}.
\end{equation}
The equation for the modes also implies the identity
\begin{equation}\label{eq:ident}
\frac{d}{dt}(\omega_{j}^{2}c_{i}c_{j}+\dot{c}_{i}\dot{c}_{j})=(\omega_{j}^{2}-\omega_{i}^{2})\dot{c}_{j}c_{i}.
\end{equation}
From the eigenfunction equation $\hat{L}e_{j}=\omega_{j}^{2}e_{j}$, we have that $-(\mu e'_{j})'=\omega_{j}^{2}\mu e_{j}$ such that
\begin{equation}\label{eq:iden1}
(\mu e'_{i}e_{j})'=(\mu e'_{i})'e_{j}+\mu e'_{i}e'_{j}=\mu(-\omega_{i}^{2}e_{i}e_{j}+e'_{i}e'_{j}).
\end{equation}
We can take the permutation $i\leftrightarrow j$ of this expression and take proper linear combinations of these two expressions to obtain the identities
\begin{equation}\label{eq:iden2}
(\omega_{j}^{2}-\omega_{i}^{2})\mu e_{j}e_{i}=(\mu(e'_{i}e_{j}-e'_{j}e_{i}))'
\end{equation}
and
\begin{equation}\label{eq:iden3}
(\omega_{j}^{2}-\omega_{i}^{2})\mu e'_{j}e'_{i}=(\mu(\omega_{j}^{2}e'_{i}e_{j}-\omega_{i}^{2}e'_{j}e_{i}))'.
\end{equation}
Subtracting (\ref{eq:A2prime}) and (\ref{eq:d2prime}), we find that
\begin{equation}
A'_{2}(x,t)-\delta'_{2}(x,t)=\frac{\nu'(x)}{\nu(x)}A_{2}(x,t)=-\nu'(x)\int_{0}^{x}\left(\dot{\phi}_{1}(y,t)^{2}+\phi'_{1}(y,t)^{2}\right)\mu(y)\text{d}y.
\end{equation}
Therefore
\begin{align*}
&\langle\left(A'_{2}-\delta'_{2}\right)\phi'_{1},e_{l}\rangle=-\int_{0}^{\frac{\pi}{2}}\text{d}x\,\phi'_{1}(x,t)e_{l}(x)\mu(x)\nu'(x)\int_{0}^{x}\text{d}y\left(\dot{\phi}_{1}(y,t)^{2}+\phi'_{1}(y,t)^{2}\right)\mu(y) \\
&=-\sum_{i=0}^{\infty}\sum_{j=0}^{\infty}\sum_{k=0}^{\infty}c_{k}(t)\int_{0}^{\frac{\pi}{2}}\text{d}x\,e'_{k}(x)e_{l}(x)\mu(x)\nu'(x) \\
&\quad\int_{0}^{x}\text{d}y\left\{\dot{c}_{i}(t)\dot{c}_{j}(t)e_{i}(y)e_{j}(y)+c_{i}(t)c_{j}(t)e'_{i}(y)e'_{j}(y)\right\}\mu(y) \\
&=-\sum_{i=0}^{\infty}\sum_{\scriptsize{\begin{matrix}j=0\\j\neq i\end{matrix}}}^{\infty}\sum_{k=0}^{\infty}\frac{c_{k}(t)}{(\omega_{j}^{2}-\omega_{i}^{2})}\int_{0}^{\frac{\pi}{2}}\text{d}x\,e'_{k}(x)e_{l}(x)(\mu(x))^{2}\nu'(x) \\
&\quad\left\{(\dot{c}_{i}(t)\dot{c}_{j}(t)+\omega_{j}^{2}c_{i}(t)c_{j}(t))e'_{i}(x)e_{j}(x)-(\dot{c}_{i}(t)\dot{c}_{j}(t)+\omega_{i}^{2}c_{i}(t)c_{j}(t))e'_{j}(x)e_{i}(x)\right\} \\
&\quad-\sum_{i=0}^{\infty}\sum_{k=0}^{\infty}c_{k}(t)\int_{0}^{\frac{\pi}{2}}\text{d}x\,e'_{k}(x)e_{l}(x)\mu(x)\nu'(x)\left\{c_{i}^{2}(t)e'_{i}(x)e_{i}(x)\mu(x)+C_{i}\int_{0}^{x}\text{d}y(e_{i}(y))^{2}\mu(y)\right\} \\
&=-2\sum_{i=0}^{\infty}\sum_{\scriptsize{\begin{matrix}j=0\\j\neq i\end{matrix}}}^{\infty}\sum_{k=0}^{\infty}\frac{c_{k}(t)H_{ijkl}}{(\omega_{j}^{2}-\omega_{i}^{2})}(\dot{c}_{i}(t)\dot{c}_{j}(t)+\omega_{j}^{2}c_{i}(t)c_{j}(t))-\sum_{i=0}^{\infty}\sum_{k=0}^{\infty}c_{k}(t)\left\{c_{i}^{2}(t)H_{iikl}+C_{i}M_{kli}\right\}.
\end{align*}
The integrals of the form $\int_{0}^{x}\text{d}y$ have been performed using the identities (\ref{eq:iden1}, \ref{eq:iden2}, \ref{eq:iden3}). Using (\ref{eq:A2}), we can proceed in a similar fashion to obtain
\begin{align*}
&\langle A_{2}\ddot{\phi}_{1},e_{l}\rangle=-\int_{0}^{\frac{\pi}{2}}\text{d}x\,\ddot{\phi}_{1}(x,t)e_{l}(x)\mu(x)\nu(x)\int_{0}^{x}\text{d}y\left(\dot{\phi}_{1}(y,t)^{2}+\phi'_{1}(y,t)^{2}\right)\mu(y) \\
&=-\sum_{i=0}^{\infty}\sum_{j=0}^{\infty}\sum_{k=0}^{\infty}\ddot{c}_{k}(t)\int_{0}^{\frac{\pi}{2}}\text{d}x\,e_{k}(x)e_{l}(x)\mu(x)\nu(x) \\
&\quad\int_{0}^{x}\text{d}y\left\{\dot{c}_{i}(t)\dot{c}_{j}(t)e_{i}(y)e_{j}(y)+c_{i}(t)c_{j}(t)e'_{i}(y)e'_{j}(y)\right\}\mu(y) \\
&=-\sum_{i=0}^{\infty}\sum_{\scriptsize{\begin{matrix}j=0\\j\neq i\end{matrix}}}^{\infty}\sum_{k=0}^{\infty}\frac{\ddot{c}_{k}(t)}{(\omega_{j}^{2}-\omega_{i}^{2})}\int_{0}^{\frac{\pi}{2}}\text{d}x\,e_{k}(x)e_{l}(x)(\mu(x))^{2}\nu(x) \\
&\quad\left\{(\dot{c}_{i}(t)\dot{c}_{j}(t)+\omega_{j}^{2}c_{i}(t)c_{j}(t))e'_{i}(x)e_{j}(x)-(\dot{c}_{i}(t)\dot{c}_{j}(t)+\omega_{i}^{2}c_{i}(t)c_{j}(t))e'_{j}(x)e_{i}(x)\right\} \\
&\quad-\sum_{i=0}^{\infty}\sum_{k=0}^{\infty}\ddot{c}_{k}(t)\int_{0}^{\frac{\pi}{2}}\text{d}x\,e_{k}(x)e_{l}(x)\mu(x)\nu(x)\left\{c_{i}^{2}(t)e'_{i}(x)e_{i}(x)\mu(x)+C_{i}\int_{0}^{x}\text{d}y(e_{i}(y))^{2}\mu(y)\right\} \\
&=2\sum_{i=0}^{\infty}\sum_{\scriptsize{\begin{matrix}j=0\\j\neq i\end{matrix}}}^{\infty}\sum_{k=0}^{\infty}\frac{\omega_{k}^{2}c_{k}(t)X_{ijkl}}{(\omega_{j}^{2}-\omega_{i}^{2})}(\dot{c}_{i}(t)\dot{c}_{j}(t)+\omega_{j}^{2}c_{i}(t)c_{j}(t))+\sum_{i=0}^{\infty}\sum_{k=0}^{\infty}\omega_{k}^{2}c_{k}(t)\left\{c_{i}^{2}(t)X_{iikl}+C_{i}W_{kli}\right\}.
\end{align*}
and
\begin{align*}
&\langle\dot{A}_{2}\dot{\phi}_{1},e_{l}\rangle=-\int_{0}^{\frac{\pi}{2}}\text{d}x\,\dot{\phi}_{1}(x,t)e_{l}(x)\mu(x)\nu(x)\int_{0}^{x}\text{d}y\frac{\partial}{\partial t}\left(\dot{\phi}_{1}(y,t)^{2}+\phi'_{1}(y,t)^{2}\right)\mu(y) \\
&=-\sum_{i=0}^{\infty}\sum_{j=0}^{\infty}\sum_{k=0}^{\infty}\dot{c}_{k}(t)\int_{0}^{\frac{\pi}{2}}\text{d}x\,e_{k}(x)e_{l}(x)\mu(x)\nu(x) \\
&\quad\int_{0}^{x}\text{d}y\frac{\partial}{\partial t}\left\{\dot{c}_{i}(t)\dot{c}_{j}(t)e_{i}(y)e_{j}(y)+c_{i}(t)c_{j}(t)e'_{i}(y)e'_{j}(y)\right\}\mu(y) \\
&=-\sum_{i=0}^{\infty}\sum_{\scriptsize{\begin{matrix}j=0\\j\neq i\end{matrix}}}^{\infty}\sum_{k=0}^{\infty}\frac{\dot{c}_{k}(t)}{(\omega_{j}^{2}-\omega_{i}^{2})}\int_{0}^{\frac{\pi}{2}}\text{d}x\,e_{k}(x)e_{l}(x)(\mu(x))^{2}\nu(x) \\
&\quad\frac{\partial}{\partial t}\left\{(\dot{c}_{i}(t)\dot{c}_{j}(t)+\omega_{j}^{2}c_{i}(t)c_{j}(t))e'_{i}(x)e_{j}(x)-(\dot{c}_{i}(t)\dot{c}_{j}(t)+\omega_{i}^{2}c_{i}(t)c_{j}(t))e'_{j}(x)e_{i}(x)\right\} \\
&\quad-\sum_{i=0}^{\infty}\sum_{k=0}^{\infty}\dot{c}_{k}(t)\int_{0}^{\frac{\pi}{2}}\text{d}x\,e_{k}(x)e_{l}(x)\mu(x)\nu(x)\frac{\partial}{\partial t}\left\{c_{i}^{2}(t)e'_{i}(x)e_{i}(x)\mu(x)+C_{i}\int_{0}^{x}\text{d}y(e_{i}(y))^{2}\mu(y)\right\} \\
&=-2\sum_{i=0}^{\infty}\sum_{j=0}^{\infty}\sum_{k=0}^{\infty}\dot{c}_{k}(t)c_{i}(t)\dot{c}_{j}(t)X_{ijkl}.
\end{align*}
In the last step, we have used the identity (\ref{eq:ident}).

From (\ref{eq:d2}), we can deduce that
\begin{align*}
&\langle\delta_{2}\ddot{\phi}_{1},e_{l}\rangle=-\int_{0}^{\frac{\pi}{2}}\text{d}x\,\ddot{\phi}_{1}(x,t)e_{l}(x)\mu(x)\int_{0}^{x}\text{d}y\left(\dot{\phi}_{1}(y,t)^{2}+\phi'_{1}(y,t)^{2}\right)\mu(y)\nu(y) \\
&=-\int_{0}^{\frac{\pi}{2}}\text{d}y\left(\dot{\phi}_{1}(y,t)^{2}+\phi'_{1}(y,t)^{2}\right)\mu(y)\nu(y)\int_{y}^{\frac{\pi}{2}}\text{d}x\,\ddot{\phi}_{1}(x,t)e_{l}(x)\mu(x) \\
&=-\sum_{k=0}^{\infty}\ddot{c}_{k}(t)\int_{0}^{\frac{\pi}{2}}\text{d}y\left(\dot{\phi}_{1}(y,t)^{2}+\phi'_{1}(y,t)^{2}\right)\mu(y)\nu(y)\left(\delta_{kl}-\int_{0}^{y}\text{d}x\,e_{k}(x)e_{l}(x)\mu(x)\right) \\
&=\sum_{\scriptsize{\begin{matrix}k=0\\k\neq l\end{matrix}}}^{\infty}\frac{\ddot{c}_{k}(t)}{(\omega_{l}^{2}-\omega_{k}^{2})}\int_{0}^{\frac{\pi}{2}}\text{d}y\left(\dot{\phi}_{1}(y,t)^{2}+\phi'_{1}(y,t)^{2}\right)(\mu(y))^{2}\nu(y)(e'_{k}(y)e_{l}(y)-e'_{l}(y)e_{k}(y)) \\
&\quad-\ddot{c}_{l}(t)\int_{0}^{\frac{\pi}{2}}\text{d}y\left(\dot{\phi}_{1}(y,t)^{2}+\phi'_{1}(y,t)^{2}\right)\mu(y)\nu(y)\left(1-\int_{0}^{y}\text{d}x(e_{l}(x))^{2}\mu(x)\right) \\
\end{align*}
\begin{align*}
&=\sum_{\scriptsize{\begin{matrix}k=0\\k\neq l\end{matrix}}}^{\infty}\sum_{i=0}^{\infty}\sum_{j=0}^{\infty}\frac{\ddot{c}_{k}(t)}{(\omega_{l}^{2}-\omega_{k}^{2})}\int_{0}^{\frac{\pi}{2}}\text{d}y\left\{\dot{c}_{i}(t)\dot{c}_{j}(t)e_{i}(y)e_{j}(y)+c_{i}(t)c_{j}(t)e'_{i}(y)e'_{j}(y)\right\} \\
&\quad(\mu(y))^{2}\nu(y)(e'_{k}(y)e_{l}(y)-e'_{l}(y)e_{k}(y)) \\
&\quad-\sum_{i=0}^{\infty}\sum_{j=0}^{\infty}\ddot{c}_{l}(t)\int_{0}^{\frac{\pi}{2}}\text{d}y\left\{\dot{c}_{i}(t)\dot{c}_{j}(t)e_{i}(y)e_{j}(y)+c_{i}(t)c_{j}(t)e'_{i}(y)e'_{j}(y)\right\} \\
&\quad\mu(y)\nu(y)\left(1-\int_{0}^{y}\text{d}x(e_{l}(x))^{2}\mu(x)\right) \\
&=-\sum_{\scriptsize{\begin{matrix}k=0\\k\neq l\end{matrix}}}^{\infty}\sum_{i=0}^{\infty}\sum_{j=0}^{\infty}\frac{\omega_{k}^{2}c_{k}(t)}{(\omega_{l}^{2}-\omega_{k}^{2})}\left\{\dot{c}_{i}(t)\dot{c}_{j}(t)(X_{klij}-X_{lkij})+c_{i}(t)c_{j}(t)(Y_{klij}-Y_{lkij})\right\} \\
&\quad+\sum_{i=0}^{\infty}\sum_{j=0}^{\infty}\omega_{l}^{2}c_{l}(t)\left\{\dot{c}_{i}(t)\dot{c}_{j}(t)P_{ijl}+c_{i}(t)c_{j}(t)B_{ijl}\right\}.
\end{align*}
We have interchanged the integration $\int\text{d}x\int\text{d}y\rightarrow\int\text{d}y\int\text{d}x$ and then separated the integral $\int_{y}^{\frac{\pi}{2}}\text{d}x=\int_{0}^{\frac{\pi}{2}}\text{d}x-\int_{0}^{y}\text{d}x$.

In a similar way, we finally obtain
\begin{align*}
&\langle\dot{\delta}_{2}\dot{\phi}_{1},e_{l}\rangle=-\int_{0}^{\frac{\pi}{2}}\text{d}x\,\dot{\phi}_{1}(x,t)e_{l}(x)\mu(x)\int_{0}^{x}\text{d}y\frac{\partial}{\partial t}\left(\dot{\phi}_{1}(y,t)^{2}+\phi'_{1}(y,t)^{2}\right)\mu(y)\nu(y) \\
&=-\int_{0}^{\frac{\pi}{2}}\text{d}y\frac{\partial}{\partial t}\left(\dot{\phi}_{1}(y,t)^{2}+\phi'_{1}(y,t)^{2}\right)\mu(y)\nu(y)\int_{y}^{\frac{\pi}{2}}\text{d}x\,\dot{\phi}_{1}(x,t)e_{l}(x)\mu(x) \\
&=-\sum_{k=0}^{\infty}\dot{c}_{k}(t)\int_{0}^{\frac{\pi}{2}}\text{d}y\frac{\partial}{\partial t}\frac{\partial}{\partial t}\left(\dot{\phi}_{1}(y,t)^{2}+\phi'_{1}(y,t)^{2}\right)\mu(y)\nu(y)\left(\delta_{kl}-\int_{0}^{y}\text{d}x\,e_{k}(x)e_{l}(x)\mu(x)\right) \\
&=\sum_{\scriptsize{\begin{matrix}k=0\\k\neq l\end{matrix}}}^{\infty}\frac{\dot{c}_{k}(t)}{(\omega_{l}^{2}-\omega_{k}^{2})}\int_{0}^{\frac{\pi}{2}}\text{d}y\frac{\partial}{\partial t}\left(\dot{\phi}_{1}(y,t)^{2}+\phi'_{1}(y,t)^{2}\right)(\mu(y))^{2}\nu(y)(e'_{k}(y)e_{l}(y)-e'_{l}(y)e_{k}(y)) \\
&\quad-\dot{c}_{l}(t)\int_{0}^{\frac{\pi}{2}}\text{d}y\frac{\partial}{\partial t}\left(\dot{\phi}_{1}(y,t)^{2}+\phi'_{1}(y,t)^{2}\right)\mu(y)\nu(y)\left(1-\int_{0}^{y}\text{d}x(e_{l}(x))^{2}\mu(x)\right) \\
\end{align*}
\begin{align*}
&=\sum_{\scriptsize{\begin{matrix}k=0\\k\neq l\end{matrix}}}^{\infty}\sum_{i=0}^{\infty}\sum_{j=0}^{\infty}\frac{\dot{c}_{k}(t)}{(\omega_{l}^{2}-\omega_{k}^{2})}\int_{0}^{\frac{\pi}{2}}\text{d}y\frac{\partial}{\partial t}\left\{\dot{c}_{i}(t)\dot{c}_{j}(t)e_{i}(y)e_{j}(y)+c_{i}(t)c_{j}(t)e'_{i}(y)e'_{j}(y)\right\} \\
&\quad(\mu(y))^{2}\nu(y)(e'_{k}(y)e_{l}(y)-e'_{l}(y)e_{k}(y)) \\
&\quad-\sum_{i=0}^{\infty}\sum_{j=0}^{\infty}\dot{c}_{l}(t)\int_{0}^{\frac{\pi}{2}}\text{d}y\frac{\partial}{\partial t}\left\{\dot{c}_{i}(t)\dot{c}_{j}(t)e_{i}(y)e_{j}(y)+c_{i}(t)c_{j}(t)e'_{i}(y)e'_{j}(y)\right\} \\
&\quad\mu(y)\nu(y)\left(1-\int_{0}^{y}\text{d}x(e_{l}(x))^{2}\mu(x)\right) \\
&=\sum_{\scriptsize{\begin{matrix}k=0\\k\neq l\end{matrix}}}^{\infty}\sum_{i=0}^{\infty}\sum_{j=0}^{\infty}\frac{\dot{c}_{k}(t)}{(\omega_{l}^{2}-\omega_{k}^{2})}\frac{\partial}{\partial t}\left\{\dot{c}_{i}(t)\dot{c}_{j}(t)(X_{klij}-X_{lkij})+c_{i}(t)c_{j}(t)(Y_{klij}-Y_{lkij})\right\} \\
&\quad-\sum_{i=0}^{\infty}\sum_{j=0}^{\infty}\dot{c}_{l}(t)\frac{\partial}{\partial t}\left\{\dot{c}_{i}(t)\dot{c}_{j}(t)P_{ijl}+c_{i}(t)c_{j}(t)B_{ijl}\right\}.
\end{align*}
Using all these expressions, we deduce from (\ref{eq:S}) the equation (\ref{eq:Sl}).

\section{FORM-based analysis of secular term coefficients}\label{sec:form}

In this section we present our FORM script dealing with the +++ secular terms, which retraces the derivation steps given in section \ref{vanish} and produces the following output, which should be matched to (\ref{Qcomp}):
\begin{verbatim}
Secular =
   + Pl*[3d+2i+2j+2k]*[(d+2i)(d+2j)(d+2k)]*[(3d+2i+2j+2k+1)eta^7-(d+i+j+k+\
   1)(3d+2i+2j+2k+2)eta^6] * ( 1 + [d+2i]^-1*[d+2j] + [d+2j]^-1*[d+2k] + 
     [d+2i]*[d+2k]^-1 );
\end{verbatim}
The script, given below, is extremely straightforward, and should be reasonably easy to interpret even for readers completely unfamiliar with FORM programming. It starts with declaring an expression matching (\ref{ppp}) up to normalization, and then applying a sequence of symbolic substitutions that implement integration by parts, re-expression through Jacobi polynomials, substituting the two highest order terms from each Jacobi polynomial, extracting the two highest-order terms of the entire (polynomial) expression, and finally, simplifying and factorizing the result. The lines starting with * are comments that do not affect the execution of the script.
\begin{verbatim}
Symbols d, Pl, eta, [eta^7], [eta^6];
Symbols i,j,k,l,any1,any2,any3;
Symbols [d+2i],[d+2j],[d+2k], [3d+2i+2j+2k], [3d+2i+2j+2k+1], [3d+2i+2j+2k+2],
         [d+2i-1], [d+2j-1], [d+2k-1],[d+i+j+k+1],[(d+2i)(d+2j)(d+2k)];
Symbols [eta^7(3d+2i+2j+2k+1)],[eta^6(d+i+j+k+1)(3d+2i+2j+2k+2)],
         [(3d+2i+2j+2k+1)eta^7-(d+i+j+k+1)(3d+2i+2j+2k+2)eta^6];
CFunctions om, H, X, Ydiff, Z, E, D, e, epr, munu, tanx, P, Ppr;

* Secular = Qijkl multiplied with (omi+omj)(omj+omk)(omi+omk)

Local Secular=
-1/2*H(i,j,k,l)*om(j)*(om(i)+om(k))*(2*om(j)+om(i)+om(k))
-1/2*H(j,k,i,l)*om(k)*(om(i)+om(j))*(2*om(k)+om(i)+om(j))
-1/2*H(k,i,j,l)*om(i)*(om(j)+om(k))*(2*om(i)+om(j)+om(k))
+X(i,j,k,l)*om(j)*om(k)*(om(j)+om(k))*
 ((om(i)+om(j))*(om(i)+om(k))+om(k)*(om(i)+om(k))+om(j)*(om(i)+om(j)))
+X(j,k,i,l)*om(i)*om(k)*(om(i)+om(k))*
 ((om(i)+om(j))*(om(j)+om(k))+om(k)*(om(j)+om(k))+om(i)*(om(i)+om(j)))
+X(k,i,j,l)*om(i)*om(j)*(om(i)+om(j))*
 ((om(i)+om(k))*(om(j)+om(k))+om(i)*(om(i)+om(k))+om(j)*(om(j)+om(k)))
-1/2*Z(i,j,k,l)*om(k)*(om(i)+om(k))*(om(j)+om(k))
-1/2*Z(j,k,i,l)*om(i)*(om(i)+om(j))*(om(i)+om(k))
-1/2*Z(k,i,j,l)*om(j)*(om(i)+om(j))*(om(j)+om(k));

id Z(i?,j?,k?,l?)=om(i)*om(j)*(X(k,l,i,j)-X(l,k,i,j))-Ydiff(k,l,i,j);

* eliminating Y

id Ydiff(i?,j?,k?,l?)=(om(i)^2-om(j)^2)*X(k,l,i,j)+om(k)^2*(X(i,j,k,l)-X(j,i,k,l))
                                        -(H(i,j,k,l)-H(j,i,k,l));
id om(l)=om(i)+om(j)+om(k);
.sort

* integrating by parts to remove the derivatives from e_l

id X(l,i?,j?,k?)=-(X(i,j,k,l)+X(j,i,k,l)+X(k,i,j,l))-E(i,j,k,l);
id H(l,i?,j?,k?)=-(H(i,k,j,l)+H(k,i,j,l))+om(j)^2*D(i,j,k,l)+4*X(j,i,k,l);

* eta=1+cos2x

id H(i?,j?,k?,l?)=epr(i)*e(j)*epr(k)*e(l)*(-d+eta);
id X(i?,j?,k?,l?)=epr(i)*e(j)*e(k)*e(l)*munu(eta);
id E(i?,j?,k?,l?)=e(i)*e(j)*e(k)*e(l)*(d-2+eta);
id D(i?,j?,k?,l?)=e(i)*e(j)*e(k)*e(l)*(-d+eta);
.sort

* substituting the mode functions with the overall normalization stripped
* and divided by (cos x)^d

id e(i?)=P(i);
id epr(i?)=-d*tanx(eta)*P(i)-4*munu(eta)*Ppr(i);

id P(l)=Pl;

id munu(eta)^2= eta*(2-eta)/4;
id tanx(eta)^2 = (2-eta)/eta;
id tanx(eta)*munu(eta)=1-(eta/2);

* 2 leading terms for each Jacobi polynomial, divided by
* eta^(i-2),  eta^(j-2),  eta^(k-2), respectively
* overall normalization stripped

id P(i)=eta^2-i*([d+2i-1]+1)*[d+2i-1]^(-1)*eta;
id P(j)=eta^2-j*([d+2j-1]+1)*[d+2j-1]^(-1)*eta;
id P(k)=eta^2-k*([d+2k-1]+1)*[d+2k-1]^(-1)*eta;

id Ppr(i)=i*eta-i*(i-1)*([d+2i-1]+1)*[d+2i-1]^(-1);
id Ppr(j)=j*eta-j*(j-1)*([d+2j-1]+1)*[d+2j-1]^(-1);
id Ppr(k)=k*eta-k*(k-1)*([d+2k-1]+1)*[d+2k-1]^(-1);

* remove low powers

id eta^i?{<6}=0;
id eta^7=[eta^7];
id eta^6=[eta^6];
.sort

id om(i)=[d+2i-1]+1;
id om(j)=[d+2j-1]+1;
id om(k)=[d+2k-1]+1;

id i=([d+2i-1]+1-d)/2;
id j=([d+2j-1]+1-d)/2;
id k=([d+2k-1]+1-d)/2;
.sort

* inverse powers of [d+2i-1],  [d+2j-1],  [d+2k-1] have cancelled out

id [d+2i-1]=[d+2i]-1;
id [d+2j-1]=[d+2j]-1;
id [d+2k-1]=[d+2k]-1;
.sort

* proceed with factorizations

id [eta^7]*[d+2i]^any1?=[eta^7]*([3d+2i+2j+2k+1]-1-[d+2j]-[d+2k])^any1;
id [3d+2i+2j+2k+1]^any1?{>1}=[3d+2i+2j+2k+1]*([d+2i]+[d+2j]+[d+2k]+1)^(any1-1);
.sort

id [eta^6]*[d+2i]^any1?=[eta^6]*([3d+2i+2j+2k+2]-2-[d+2j]-[d+2k])^any1;
id [3d+2i+2j+2k+2]^any1?{>1}=[3d+2i+2j+2k+2]*([d+2i]+[d+2j]+[d+2k]+2)^(any1-1);
.sort

id [eta^6]*[d+2i]^any1?=[eta^6]*(2*[d+i+j+k+1]-2+d-[d+2j]-[d+2k])^any1;
id [d+i+j+k+1]^any1?{>1}=[d+i+j+k+1]*(([d+2i]+[d+2j]+[d+2k]+2-d)/2)^(any1-1);
.sort

id [eta^6]*[d+i+j+k+1]*[3d+2i+2j+2k+2]=[eta^6(d+i+j+k+1)(3d+2i+2j+2k+2)];
id [eta^7]*[3d+2i+2j+2k+1]=[eta^7(3d+2i+2j+2k+1)];
.sort

id [eta^7(3d+2i+2j+2k+1)]=[(3d+2i+2j+2k+1)eta^7-(d+i+j+k+1)(3d+2i+2j+2k+2)eta^6]
                                        +[eta^6(d+i+j+k+1)(3d+2i+2j+2k+2)];
.sort

id [d+2k]=[3d+2i+2j+2k]-[d+2i]-[d+2j];
id [3d+2i+2j+2k]^any1?{>1}=[3d+2i+2j+2k]*([d+2i]+[d+2j]+[d+2k])^(any1-1);
.sort

id Pl=Pl*[(d+2i)(d+2j)(d+2k)]*[d+2i]^(-1)*[d+2j]^(-1)*[d+2k]^(-1);

Bracket Pl,[(3d+2i+2j+2k+1)eta^7-(d+i+j+k+1)(3d+2i+2j+2k+2)eta^6],
               [3d+2i+2j+2k],[(d+2i)(d+2j)(d+2k)];

Print;
.end
\end{verbatim}

(It may appear surprising that the output of our script is not manifestly symmetric under all permutations of $i$, $j$ and $k$, even if the starting expression is. There is no contradiction here, however, since the integration by parts (\ref{Yint}) we employ only preserves the permutation symmetry of the total integral expressions, but upsets some of the permutation symmetries of the integrands. In our case, since the end result of the integration is 0, it is trivially fully permutation-symmetric, even if an explicit formula for the integrand we give is not.)

We have also employed a very similar script, with the roles played by $i$ and $l$ interchanged and starting with (\ref{ppm}) rather than (\ref{ppp}), to analyze the +\,-\,- secular terms and derive the coefficients of the two highest powers in ${\cal U}_{jkl}(\xi)$ of (\ref{Upoly}), which can be read off from the following output (what matters for us is the ratio of the two coefficients, hence only the expression in the square brackets of the first line of the output is relevant):
\begin{verbatim}
 Secular =
    + Pi*[(3d+2l+2j+2k+1)eta^7-(d+l+j+k+1)(3d+2l+2j+2k+2)eta^6] * ( 2*
     [d+2j]*[d+2k]^3 + 2*[d+2j]^2*[d+2k]^2 + 2*[d+2j]^3*[d+2k] + [d+2l]*
     [d+2k]^3 + [d+2l]*[d+2j]*[d+2k]^2 + 2*[d+2l]*[d+2j]^2*[d+2k] + [d+2l]
     *[d+2j]^3 - [d+2l]^2*[d+2k]^2 - [d+2l]^2*[d+2j]*[d+2k] + 2*[d+2l]^2*
     [d+2j]^2 + [d+2l]^3*[d+2j] );
\end{verbatim}



\begin{thebibliography}{99}

\bibitem{landau}L.~D.~Landau and E.~.M.~Lifshitz, {\it Mechanics}, 3rd edition, Reed Publishing (1981),  pp. 86-87.

\bibitem{logan}J.~D.~Logan. {\it Applied Mathematics}, 3rd edition, John Wiley \& Sons (2006) pp. 93-94.

\bibitem{Chen:1995ena}
  L.~-Y.~Chen, N.~Goldenfeld and Y.~Oono,
 {\em ``The Renormalization group and singular perturbations: Multiple scales, boundary layers and reductive perturbation theory,''}
  Phys.\ Rev.\ E {\bf 54} (1996) 376
  \arXiv{arXiv:hep-th/9506161}.

\bibitem{KAM} V.~I.~Arnol'd, V.~V.~Kozlov and A.~I.~Neistadt,
{\it Mathematical aspects of classical and celestial mechanics}, Springer (1997).

\bibitem{Nakayama:2013fha}
  Y.~Nakayama,
  {\em ``Holographic interpretation of renormalization group approach to singular perturbations in nonlinear differential equations,''}
  Phys.\ Rev.\ D {\bf 88} (2013) 10,  105006
  \arXiv{arXiv:1305.4117} [hep-th].

\bibitem{Kuperstein:2013hqa}
  S.~Kuperstein and A.~Mukhopadhyay,
  {\em ``Spacetime emergence via holographic RG flow from incompressible Navier-Stokes at the horizon,''}
  JHEP {\bf 1311} (2013) 086
  \arXiv{arXiv:1307.1367} [hep-th].

\bibitem{Bizon:2011gg}
  P.~Bizon and A.~Rostworowski,
 {\em ``On weakly turbulent instability of anti-de Sitter space,''}
  Phys.\ Rev.\ Lett.\  {\bf 107} (2011) 031102
  \arXiv{arXiv:1104.3702} [gr-qc].

\bibitem{Dias:2011ss}
  O.~J.~C.~Dias, G.~T.~Horowitz and J.~E.~Santos,
  {\em ``Gravitational Turbulent Instability of Anti-de Sitter Space,''}
  Class.\ Quant.\ Grav.\  {\bf 29} (2012) 194002
  \arXiv{arXiv:1109.1825} [hep-th].

\bibitem{Dias:2012tq}
  O.~J.~C.~Dias, G.~T.~Horowitz, D.~Marolf and J.~E.~Santos,
  {\em``On the Nonlinear Stability of Asymptotically Anti-de Sitter Solutions,''}
  Class.\ Quant.\ Grav.\  {\bf 29} (2012) 235019
  \arXiv{arXiv:1208.5772} [gr-qc].

\bibitem{Maliborski:2013jca}
  M.~Maliborski and A.~Rostworowski,
 {\em ``Time-Periodic Solutions in an Einstein AdS-Massless-Scalar-Field System,''}
  Phys.\ Rev.\ Lett.\  {\bf 111} (2013) 5,  051102
  \arXiv{arXiv:1303.3186} [gr-qc].

\bibitem{Buchel:2013uba}
  A.~Buchel, S.~L.~Liebling and L.~Lehner,
{\em ``Boson stars in AdS spacetime,''}
  Phys.\ Rev.\ D {\bf 87} (2013) 12,  123006
  \arXiv{arXiv:1304.4166} [gr-qc].

\bibitem{Abajo-Arrastia:2014fma}
  J.~Abajo-Arrastia, E.~da Silva, E.~Lopez, J.~Mas and A.~Serantes,
  {\em ``Holographic Relaxation of Finite Size Isolated Quantum Systems,''}
  \arXiv{arXiv:1403.2632} [hep-th].

\bibitem{Maliborski:2014rma}
  M.~Maliborski and A.~Rostworowski,
  {\em``What drives AdS unstable?,''}
  \arXiv{arXiv:1403.5434} [gr-qc].

\bibitem{Balasubramanian:2014cja}
  V.~Balasubramanian, A.~Buchel, S.~R.~Green, L.~Lehner and S.~L.~Liebling,
 {\em ``Holographic Thermalization, Stability of AdS, and the FPU Paradox,''}
  \arXiv{arXiv:1403.6471} [hep-th].

\bibitem{murdock}J.~A.~Murdock, {\it Perturbations: Theory and Methods}, SIAM (1987).

\bibitem{shentangwang}J.~Shen, T.~Tang, L.-L.~Wang, {\it  Spectral Methods: Algorithms, Analysis and Applications}, Springer (2011).

\bibitem{FORM} J.~Kuipers, T.~Ueda, J.~A.~M.~Vermaseren and J.~Vollinga,
  {\em ``FORM version 4.0,''}
  Comput.\ Phys.\ Commun.\  {\bf 184} (2013) 1453
  \arXiv{arXiv:1203.6543} [cs.SC].

\end{thebibliography}
\end{document}